\documentclass[twocolumn,aps,pre,superscriptaddress]{revtex4-2}
\usepackage[utf8]{inputenc}
\usepackage{amsmath}
\usepackage{amssymb}
\usepackage{amscd}
\usepackage{graphicx}
\usepackage{bbm}
\usepackage{dsfont}
\usepackage[usenames,dvipsnames,svgnames,table]{xcolor}
\usepackage{datetime}
\usepackage{physics}
\usepackage{comment}
\usepackage{cancel}
\usepackage{float}
\usepackage{ulem}
\usepackage{flushend}
\usepackage{relsize}
\usepackage{siunitx}
\longdate

\begin{document}

\title{Three-Dimensional Spontaneous Flow Transition in a Homeotropic Active Nematic}
\author{Vincenzo J. Pratley}
\email{vj.pratley@warwick.ac.uk}
\affiliation{Department of Physics, Gibbet Hill Road, University of Warwick, Coventry, CV4 7AL, United Kingdom.}
\author{Enej Caf}
\email{enej.caf@fmf.uni-lj.si}
\affiliation{Faculty of Mathematics and Physics, University of Ljubljana, Jadranska 19, SI-1000 Ljubljana, Slovenia.}
\author{Miha Ravnik}
\email{miha.ravnik@fmf.uni-lj.si}
\affiliation{Faculty of Mathematics and Physics, University of Ljubljana, Jadranska 19, SI-1000 Ljubljana, Slovenia.}
\affiliation{Condensed Matter Department, J. Stefan Institute, Jamova 39, SI-1000 Ljubljana, Slovenia.}
\author{Gareth P. Alexander}
\email{g.p.alexander@warwick.ac.uk}
\affiliation{Department of Physics, Gibbet Hill Road, University of Warwick, Coventry, CV4 7AL, United Kingdom.}

\date{\today}

\begin{abstract}
We study the three-dimensional spontaneous flow transition of an active nematic in an infinite slab geometry using a combination of numerics and analytics. We show that it is determined by the interplay of two eigenmodes -- called S- and D-mode -- that are unstable at the same activity threshold and spontaneously breaks both rotational symmetry and chiral symmetry. The onset of the unstable modes is described by a non-Hermitian integro-differential operator, which we determine their exponential growth rates from using perturbation theory. The S-mode is the fastest growing. After it reaches a finite amplitude, the growth of the D-mode is anisotropic, being promoted perpendicular to the S-mode and suppressed parallel to it, forming a steady state with a full three-dimensional director field and a well-defined chirality. Lastly, we derive a model of the leading-order time evolution of the system close to the activity threshold.
\end{abstract}
\maketitle

\section{Introduction}
\label{sec:introduction}

Active matter is a class of materials that lie outside of thermodynamic equilibrium due to the conversion of energy consumed by the constituent particles to mechanical work \cite{RamaswarmyMech2010,marchetti2013hydrodynamics,finlayson1969}. Active matter can be considered an active nematic whenever when the constituent particles display orientational order akin to a nematic liquid crystal~\cite{doostmohammadi2018active}. Such systems can be natural, such as cell colonies~\cite{duclos2017,meacock2021,copenhagen2021}, epithelial tissues~\cite{saw2017topological,doostmohammadi2022}, bacterial suspensions~\cite{wensink2012meso,wioland2016,dunkel2013fluid}, and microtubule and motor protein mixtures~\cite{sanchez2012spontaneous}, or artificial, such as vibrated granular rods~\cite{narayan2007,kumar2014flocking}. A key property of active matter is the emergence of spontaneous, collective motion on scales much larger that that of the individual constituents. This has important real-world implications. In biology, for example, collective motion plays a role during organ formation and development~\cite{mclennan2012multiscale} and wound healing~\cite{poujade2007collective}. There is also potential to harness the self-generated flows of active nematic materials to create self-operating microfluidic devices that do not rely on external forcing, or to incorporate other aspects of passive liquid crystals, such as utilising colloidal inclusions~\cite{houston2023,houston2023cog,ray2023}.

Active nematic systems may be modelled by adapting the well-established dynamical equations of passive nematic liquid crystals~\cite{beris1994thermodynamics,deGennesProst} to include active terms~\cite{doostmohammadi2018active,simha2002hydrodynamic}. One key triumph of the theory of active nematics is the prediction that such systems will spontaneously transition to a flowing state on their own accord due to their fundamental hydrodynamic instability~\cite{simha2002hydrodynamic}. In unbounded systems, this instability sets in at arbitrarily long perturbation wavelengths and the system eventually transitions to a chaotic state known as active turbulence~\cite{doostmohammadi2018active,sanchez2012spontaneous,wensink2012meso,dunkel2013fluid,alert2022active}. Confinement of active nematic systems can suppress the onset of active turbulence and instead the hydrodynamic instability acts to produce non-chaotic flows, first predicted theoretically by Voituriez \textit{et al.}~\cite{voituriez2005} and later confirmed in simulations performed by Marenduzzo \textit{et al.}~\cite{marenduzzo2007}. The spontaneous flow transition has been observed in experiments on spindle-shaped cells in confined strips~\cite{duclos2018}, demonstrating potential relevance to cell transport in development or cancer. 

The confinement of active nematics and the resulting spontaneous flows have been a topic of great interest to the scientific community~\cite{thampi2022channel}. Most research has focused on two-dimensional systems~\cite{wioland2016,zumdieck2008,edwards2009,furthauer2012,ravnik2013,shendruk2017dancing,doostmohammadi2017,chen2018dynamics,hardouin2019reconfigurable,opathalage2019self,chandragiri2019active,samui2021flow,rorai2021active}, but more recently the attention has shifted towards understanding three-dimensional systems~\cite{wu2017transition,shendruk2018twist,chandragiri2020,chandrarkar2020,varghese2020,strubing2020,fan2021effects,keogh2022helical}.

Of particular interest to us are the spontaneous flow transitions within rectangular channels. Different flow states can be found, depending the boundary conditions and parameters. Flows can be roughly separated into two categories: streaming flow states and swirling flow states~\cite{thampi2022channel}. These two categories can be further sub-divided. For example, the streaming flow category can be subdivided into Poiseuille-like flows~\cite{samui2021flow,chandragiri2020}, shear-like flows~\cite{duclos2018}, oscillatory flows~\cite{samui2021flow,chandragiri2020}, grinder train flows and double helix-like flows~\cite{keogh2022helical}. The latter two are only seen in three dimensions and possess non-zero helicity and are yet to be seen experimentally. 

\begin{figure*}[t]
\centering
\includegraphics[width=\linewidth]{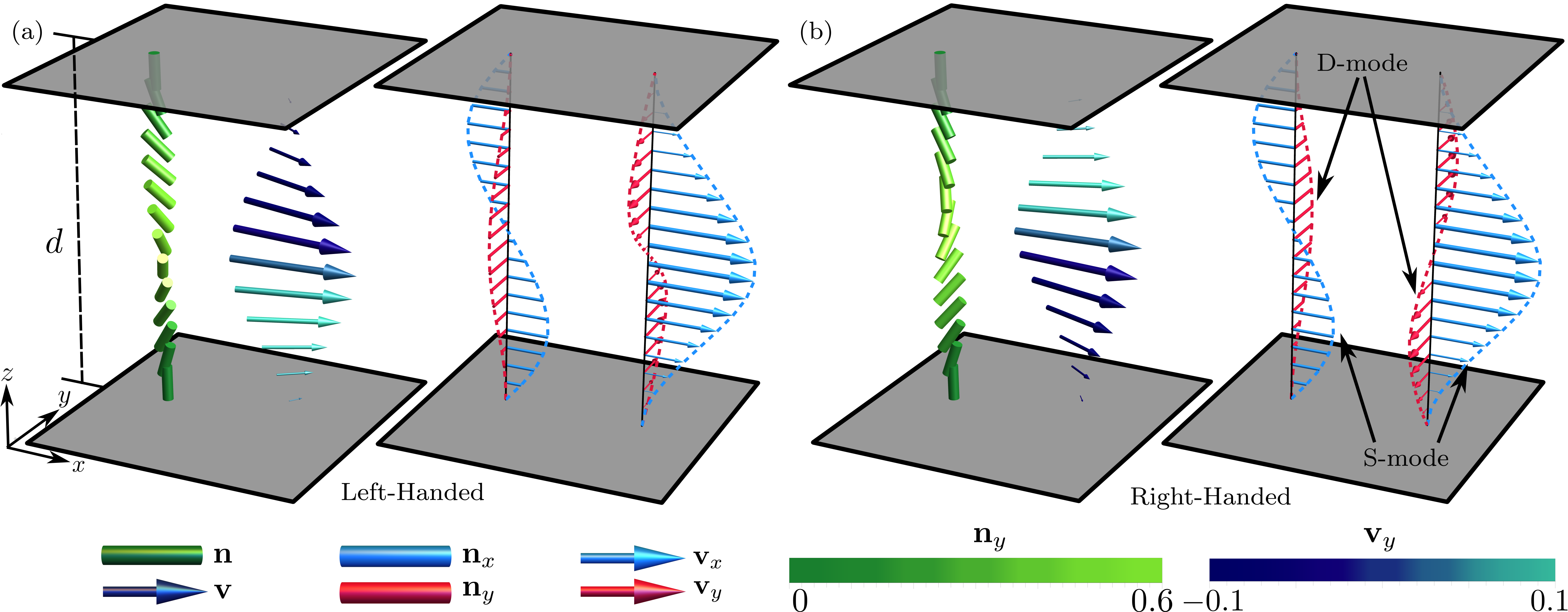}
\caption{Director and flow fields of the spontaneous flow transition. (a) Left: steady-state director and velocity fields after the spontaneous flow transition with left-handed chirality. Right: decomposition into the $x$ and $y$ components of the left-handed steady-state director and velocity fields. (b) Left: steady-state director and velocity fields after the spontaneous flow transition with right-handed chirality. Right: decomposition into the $x$ and $y$ components of the right-handed steady-state director and velocity fields.}
\label{fig:schematic}
\end{figure*}

Here, we study the spontaneous flow transition for an active nematic in a three-dimensional cell with normal anchoring boundary conditions. This geometry is analogous to the Frederiks transition in a homeotropic cell. We find that the transition leads to a twisted director field and a spontaneous flow that has both Poiseuille-like and shear-like components. The twist is right-handed or left-handed with equal probability and represents a spontaneous chiral symmetry breaking, in addition to the spontaneous rotational symmetry breaking of the direction of the Poiseuille-like flow. We identify the reason for this as the degeneracy of two eigenmodes of the linear stability operator for the system at the threshold of instability. We believe that this is an accidental degeneracy, rather than arising due to some underlying symmetry. We label these modes the S-mode and the D-mode. The degeneracy is accidental, rather than arising from an underlying symmetry, and clarifies some aspects of the existing literature for planar anchoring. 
We develop a hierarchical perturbative analysis of the growth of both modes above threshold that reproduces all aspects of the instability in excellent agreement with full numerical simulations.

\section{Spontaneous Flow Transition}
\label{sec:flow_state}

We consider an extensile, uniaxial active nematic confined between two infinite, parallel plates with a fixed cell gap, $d$. We assume no slip boundary conditions and strong homeotropic anchoring on both plates. For this anchoring condition and with the normal of the plates being $\vb{e}_z$, the ground state (i.e. the state that the system is in below threshold) director field is $\vb{n}=\vb{e}_z$ which possesses evident rotational symmetry around the $z$ axis. The setup is shown in  Fig.~\ref{fig:schematic}.

We establish the basic character of the active instability and spontaneous flow transition by performing numerical simulations with random initial perturbations to the ground state. We find that there is a threshold in activity, below which the system remains in the ground state and above which the system spontaneously starts flowing. The flow field consists of a Poiseuille-like component and a shear-like component perpendicular to it. The Poiseuille-like flow component results in a net flux within the system, the direction of which is random and represents spontaneous rotational symmetry breaking. The director field is twisted with either a right or left handedness, occurring with equal probability. Hence, the system also undergoes spontaneous chiral symmetry breaking. We note that the shear-like component of the flow is reversed between the two possible twist configurations. The director and flow fields are shown in Fig.~\ref{fig:schematic}.

This twisted flow state arises from the coupled evolution of two degenerate eigenmodes that both become unstable at the activity threshold. We believe that this is an accidental degeneracy, rather than arising due to some underlying symmetry. The degeneracy may be lifted by applying a generic perturbation, such as the application of an electric field.  We label these modes the S-mode and the D-mode. The different chiralities emerge from the fact that the D-mode can evolve in one of two possible directions perpendicular to the S-mode, with each direction being equally probable. The flow components associated with the S-mode and D-mode are the Poiseuille-like and shear-like flows respectively.

\section{Linear Instability and Threshold}
\label{sec:linear_threshold}

\subsection{Active Nematic Hydrodynamics}

Active nematic systems can be modelled by the active Beris-Edwards equations~\cite{doostmohammadi2018active,beris1994thermodynamics}
\begin{gather}
    \partial_t \rho + \div \bigl( \rho \vb{v}\bigr)=0,\label{eqn:continuityN} \\
    \rho \partial_t \vb{v} + \rho \vb{v}\vdot \grad \vb{v}= \div \vb*{\Pi}, \label{eqn:Navier-StokesN} \\
    \left(\partial_t+\vb{v}\vdot\grad \right) \vb{Q}=\Gamma \vb{H} + \vb{S},
    \label{eqn:Beris-Edwards}
\end{gather} 
which describe the coupled evolution of the fluid density, $\rho$, velocity, $\vb{v}$, and the nematic order paramerter, $\vb{Q}$.  We solve the Beris-Edwards equations numerically using a hybrid lattice Boltzmann algorithm~\cite{kruger2017lattice}, with full details given in \S\ref{sec:methods}. The activity is incorporated into~\eqref{eqn:Navier-StokesN} in the usual way by adding an additional contribution to the stress,  $\vb*{\Pi}^{\textrm{a}}=-\zeta_\textrm{LB} \vb{Q}$, modelling a force dipole at the microscopic level with a strength given by the phenomenological activity parameter, $\zeta_\textrm{LB}$. Extensile activity corresponds to $\zeta_\textrm{LB}>0$ and contractile activity to $\zeta_\textrm{LB}<0$.

In the analytical analysis, we work in terms of the director field, reducing the Beris-Edwards nematodynamic equations to the Ericksen-Leslie form~\cite{marenduzzo2007}. Assuming low Reynolds number, constant density, and a uniaxial form for the nematic order parameter, $Q_{ij}=\frac{3S}{2}\bigl(n_in_j-\delta_{ij}/3\bigr)$, with constant $S$, one writes:
\begin{gather}
    \div\vb{v} = 0 , \label{eqn:continuity}\\
    - \grad p + \mu \nabla^2 {\bf v} + \div \bigl( \vb*{\sigma}^{\textrm{el}} + \vb*{\sigma}^{\textrm{a}} \bigr) = 0, \label{eqn:Stokes_n}\\
    \partial_t\vb{n}+\vb{v}\vdot\grad\vb{n}+\vb*{\Omega}\vb{n}=\frac{1}{\gamma}\vb{h}-\nu\Bigl[\vb{D}\vb{n}-\bigl(\vb{n}\vdot\vb{D}\vb{n}\bigr)\vb{n}\Bigr] . \label{eqn:n_time_evolution}
\end{gather}
In~\eqref{eqn:Stokes_n}, $\vb*{\sigma}^{\textrm{el}}$ denotes the elastic stresses coming from the nematic director and the active stress is $\boldsymbol{\sigma}^{\textrm{a}} = -\zeta {\bf nn}$, where $\zeta=\frac{3S}{2}\zeta_\textrm{LB}$. We consider only the flow aligning regime, where the flow aligning parameter $\nu<-1$~\cite{deGennesProst}. Further relevant aspects of the correspondence between the Beris-Edwards and Ericksen-Leslie equations are given in \S\ref{sec:methods}.

\subsection{Linear Stability Analysis}

We start by considering the Ericksen-Leslie formalism in quasi-one-dimensional geometry where the spatial dependence is only along the cell normal ($z$-direction) but the flow field, $\vb{v}$, and active nematic director, $\vb{n}$, can be in any 3D direction. The continuity equation then implies that $v_z = 0$ and the Stokes equation~\eqref{eqn:Stokes_n} can be integrated directly to give   
\begin{gather}
    p = \sigma_{zz}^\textrm{el} + \sigma_{zz}^\textrm{a} + \textrm{constant}, \label{eqn:pressure_uniform_soln}\\
    v_i = \frac{1}{\mu} \biggl( z \bigl\langle \sigma_{iz}^{\textrm{el}} + \sigma_{iz}^{\textrm{a}} \bigr\rangle -\int_0^z \sigma_{iz}^{\textrm{el}} + \sigma_{iz}^{\textrm{a}} \,\dd u \biggr). \label{eqn:velocity_uniform_soln}
\end{gather}
Here, the notation $\langle\cdots\rangle=\frac{1}{d}\int_0^d \cdots \,\dd u$ represents the average of the argument over the cell gap; these terms arise from the no slip condition at the two cell boundaries, $z = 0, d$. For the director dynamics, we will find it convenient to write ${\bf n}$ in the form 
\begin{equation}
    \vb{n} = \cos\varphi\bigl(\cos\theta\,\vb{e}_z+\sin\theta\,\vb{e}_x\bigr)+\sin\varphi\,\vb{e}_y , \label{eqn:3D_director}
\end{equation}
parameterised by two angles $\theta$ and $\varphi$, in terms of which the director dynamics~\eqref{eqn:n_time_evolution} becomes   
\begin{gather}
    \partial_{t} \theta  = \frac{1}{\cos\varphi} \,\vb{m}_\theta \vdot \biggl[ \frac{1}{\gamma} \,\vb{h} - \bigl( \vb*{\Omega} \vb{n} + \nu \,\vb{D}\vb{n} \bigr) \biggr] , \label{eqn:theta_S_time_vector_form} \\
    \partial_{t} \varphi  = \vb{m}_\varphi \vdot \biggl[ \frac{1}{\gamma} \,\vb{h} - \bigl( \vb*{\Omega} \vb{n} + \nu \,\vb{D}\vb{n} \bigr) \biggr] , \label{eqn:theta_D_time_vector_form}
\end{gather}
where we have defined the unit vectors 
\begin{gather}
    \vb{m}_\theta = -\sin\theta\,\vb{e}_z+\cos\theta\,\vb{e}_x,\label{eqn:m_s_vector}\\
    \vb{m}_\varphi = -\sin\varphi\bigl(\cos\theta\,\vb{e}_z+\sin\theta\,\vb{e}_x\bigr)+\cos\varphi\,\vb{e}_y. \label{eqn:m_D_vector}
\end{gather}
Substituting the flow solution~\eqref{eqn:velocity_uniform_soln} for ${\bf D}$ and $\boldsymbol{\Omega}$, \eqref{eqn:theta_S_time_vector_form} and~\eqref{eqn:theta_D_time_vector_form} reduce to a pair of coupled, nonlinear integro-differential equations for the two angles, which we give in full in Appendix~\ref{app:full_evolution_equations}. Both equations have the same linearisation, which we write only for $\theta$, 
\begin{equation}
\begin{split}
    \partial_t \theta & = \frac{K}{\gamma} \partial_{z}^2 \theta + \frac{K(1-\nu)^2}{4\mu} \Bigl( \partial_{z}^2 \theta - \bigl\langle \partial_{z}^2 \theta \bigr\rangle \Bigr)\\
    & \quad + \frac{\zeta(1-\nu)}{2\mu} \Bigl( \theta - \bigl\langle\theta \bigr\rangle \Bigr) \equiv {\cal L}\, \theta .
\end{split}
\label{eqn:linear_theta_operator}
\end{equation}
The uniform state ($\theta = 0$) is linearly unstable when the linear integro-differential operator, ${\cal L}$, has a positive eigenvalue, $\lambda$. A perturbation along the associated eigenfunction then grows exponentially with rate $\lambda$ until it saturates at a steady state solution of the full nonlinear equations. The eigenfunctions of ${\cal L}$ separate into two symmetry classes according to whether they are odd or even about the cell midplane. 

\begin{figure}[tb]
\centering
{\includegraphics[width=0.5\textwidth]{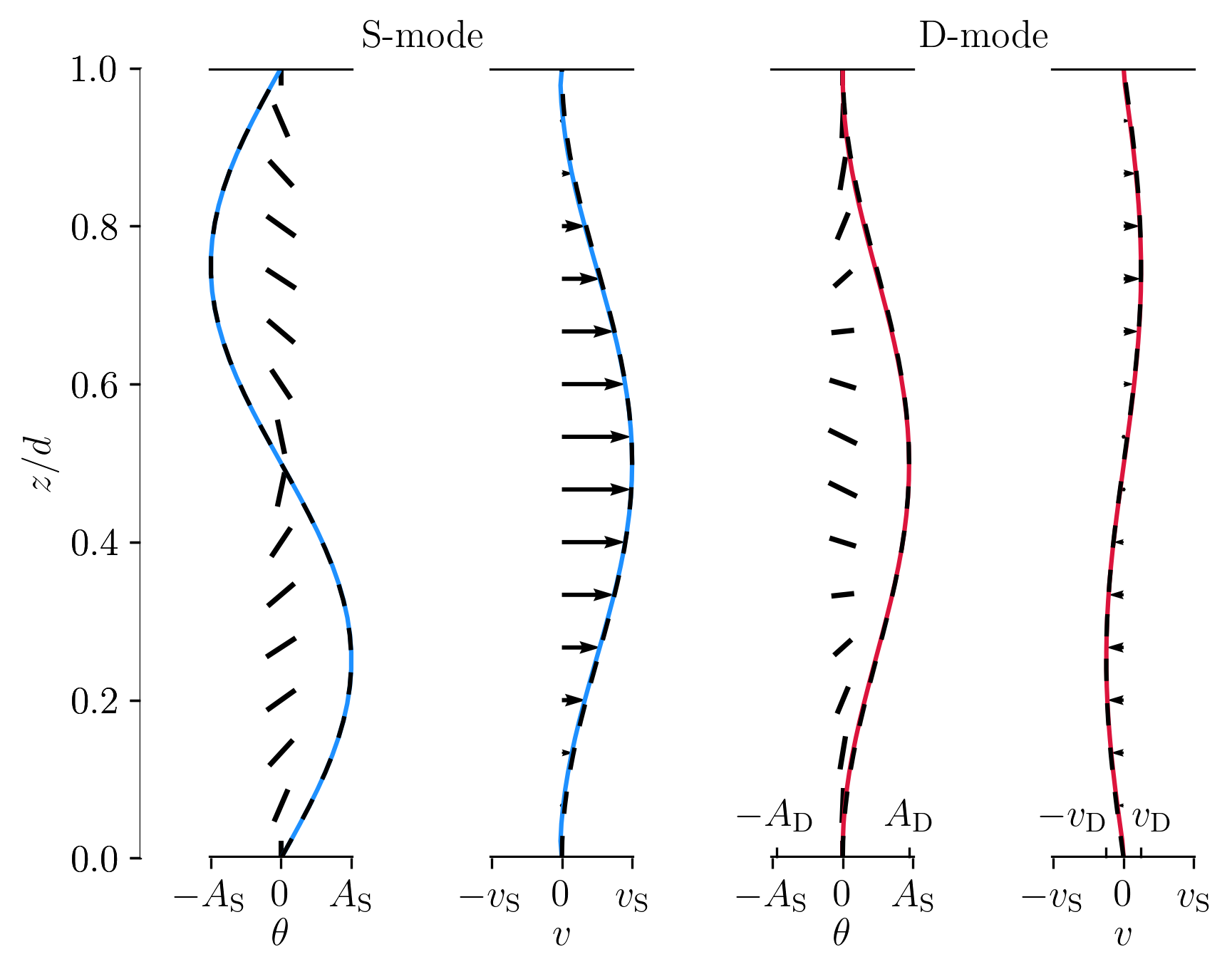}}
\caption{Comparison of the simulation results of the director and flow profiles (solid blue and red curves) with the corresponding analytical predictions (black dashed curves) for activity in the vicinity of the threshold activity, $\zeta_\textrm{LB}=0.063$. The blue curves show the director (left) and velocity (right) profiles of the S-mode, so called because of its `S'-like appearance across the cell gap. The red curves show the director (left) and velocity (right) profiles of the D-mode, again, named after the `D'-like appearance of its director profile. The observed amplitude of the S-mode profile is $A_\mathrm{S}= 0.0043$ and the velocity amplitude is $v_\mathrm{S}=1.71  \cdot 10^{-4} \,\,\, \Gamma L/\xi_n$. Observed ratios between S- and D-mode are $A_\textrm{D}/A_\textrm{S}=0.95$ and $v_{\mathrm{D}}/v_{\mathrm{S}}= 0.25$.
}
\label{fig:2dprofile}
\end{figure}

For the odd eigenfunctions, the integral terms in~\eqref{eqn:linear_theta_operator} vanish and ${\cal L}$ reduces to a Schr\"odinger-type operator whose eigenfunctions are 
\begin{equation}
    \theta = A_{\textrm{S}} \sin\frac{2n\pi z}{d} , \quad n \in \mathbb{N} ,
    \label{eqn:theta_S_linear}
\end{equation}
where $A_{\textrm{S}}$ is an amplitude. The lowest mode, $n=1$, becomes unstable first, which happens at the threshold activity 
\begin{equation}
    \zeta_{\textrm{th}} = \frac{8\pi^2\mu K}{\gamma (1-\nu) d^2} \biggl[ 1 + \frac{\gamma(1-\nu)^2}{4\mu} \biggr] .
\label{eqn:activity_threshold}
\end{equation}
In the flow aligning regime ($\nu < -1$), which we restrict our attention to, the instability is for extensile activity ($\zeta > 0$). This unstable mode is associated with a flow 
\begin{equation}
    v = \frac{4\pi K A_\textrm{S}}{\gamma(1-\nu)d} \biggl( 1 -\cos\frac{2\pi z}{d} \biggr) ,
    \label{eqn:s-mode_vel}
\end{equation}
that is even about the cell midplane and represents a fluid flux along a spontaneously chosen direction. We refer to this unstable mode, and the steady spontaneous flow state it evolves into, as the `S-mode' due to the appearance of the director across the cell gap. 

For the even eigenfunctions of ${\cal L}$, the integral terms in \eqref{eqn:linear_theta_operator} do not vanish and we have not found closed-form expressions for all of the eigenfunctions. However, one can verify directly that 
\begin{equation}
    \theta = \frac{A_{\textrm{D}}}{2} \biggl( 1 - \cos \frac{2\pi z}{d} \biggr) ,
    \label{eqn:theta_D_linear}
\end{equation}
is an eigenfunction with eigenvalue zero at the threshold activity, $\zeta = \zeta_{\textrm{th}}$. $A_{\textrm{D}}$ is an amplitude for the mode. The associated fluid flow  
\begin{equation}
    v = -  \frac{2\pi K A_\textrm{D}}{\gamma(1-\nu)d} \sin\frac{2\pi z}{d},
    \label{eqn:D-mode_vel}
\end{equation}
is shear-like and odd about the cell midplane with no net flux. As before, the direction is chosen spontaneously. We refer to this unstable mode as the `D-mode', again due to the appearance of the director across the cell gap. 

Numerically, we seeded an S-mode of the form~\eqref{eqn:theta_S_linear} and a D-mode of the form~\eqref{eqn:theta_D_linear} separately and letting them evolve into steady state for activities very close to the threshold. To do this, we reduced our simulation box to 3 bulk points in the direction perpendicular to the flow, effectively reducing the system to two dimensions so as to more easily isolate the individual eigenmodes. The results of the S- and D-modes are shown with their associated flow fields in Fig.~\ref{fig:2dprofile}, along with direct comparison to analytical predictions, where there is excellent agreement.

Overall, the spontaneous flow instability with homeotropic and no-slip boundary conditions is characterised by having two degenerate modes, one in each symmetry class, that become linearly unstable at the same threshold activity, each with a spontaneously chosen in-plane direction. This degeneracy in the linear instability distinguishes the active spontaneous flow transition from the Frederiks transition in a passive system, where the instability is to the fundamental mode in the even sector at a threshold well below that of the first mode in the odd sector~\cite{deGennesProst}.

\section{Growth Rates Above Threshold}
\label{sec:evol_above_thresh}

Above the threshold activity both unstable modes will grow exponentially at rates given by their respective eigenvalues of the linear stability operator ${\cal L}$. We first determine these for the S- and D-modes separately and subsequently consider how they coevolve. The S-mode~\eqref{eqn:theta_S_linear} is an eigenfunction of ${\cal L}$ for all values of the activity, with eigenvalue 
\begin{equation}
    \lambda_\textrm{S} = \frac{1-\nu}{2\mu} \bigl( \zeta - \zeta_{\textrm{th}} \bigr) . 
    \label{eqn:S_mode_eigenvalue}
\end{equation}
For the D-mode, the expression~\eqref{eqn:theta_D_linear} is an exact eigenfunction only at the threshold activity, $\zeta = \zeta_{\textrm{th}}$, where the eigenvalue is zero. We do not have its closed form more generally. However, for activities close to threshold we can determine the eigenvalue from perturbation theory, expanding to first order in $\zeta-\zeta_{\textrm{th}}$. We provide the details in \S\ref{sec:methods} and state here only the result 
\begin{equation}
    \lambda_\textrm{D} = \frac{1-\nu}{6\mu+\gamma(1-\nu)^2}\bigl( \zeta - \zeta_{\textrm{th}} \bigr) .
    \label{eqn:D_mode_eigenvalue}
\end{equation}
Comparing against the eigenvalue of the S-mode gives a ratio $\lambda_{\textrm{S}}/\lambda_{\textrm{D}} = 3 + \frac{\gamma (1-\nu)^2}{2\mu}$, from which we see that the S-mode will grow fastest above threshold. As a result, unless it is suppressed, it will dominate the initial evolution of the system post instability. 

\begin{figure}[t]
\includegraphics[width=0.5 \textwidth]{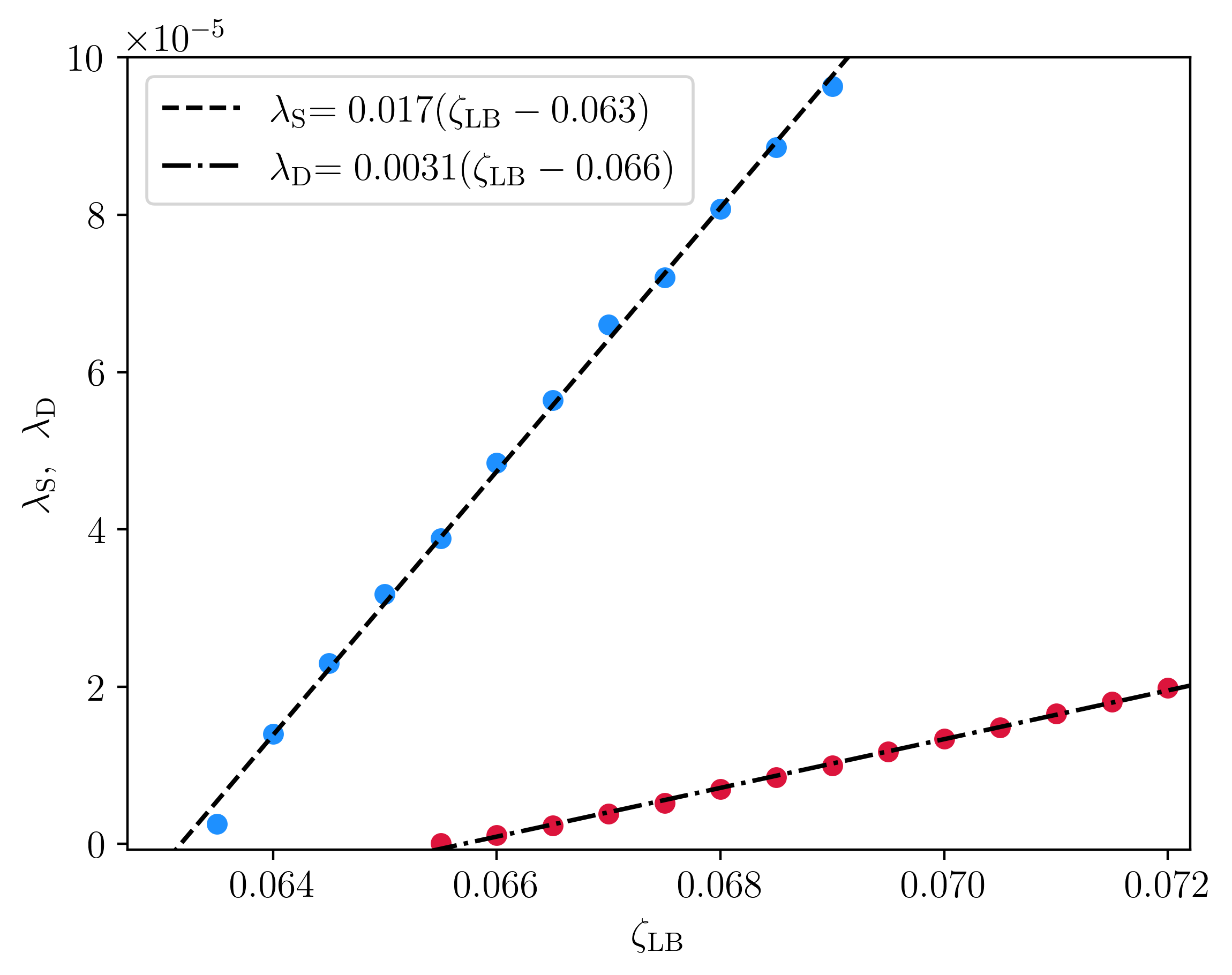}
\caption{Exponential growth rates vs activity for an individually seeded S-mode (blue), and D-mode (red). The black dashed lines are linear fits of the numerical data  close to threshold, where the growth rates exhibit a linear relation with activity.}
\label{fig:growth-rates-single}
\end{figure}

To allow for comparison with numerical simulations, we convert \eqref{eqn:S_mode_eigenvalue} and \eqref{eqn:D_mode_eigenvalue} to simulation units.
\begin{align}
    \lambda_\textrm{S} &=0.0178 \bigl(\zeta_\textrm{LB} - 0.0629\bigr),\label{eqn:S_mode_eigenvalue_numerical_units}\\
    \lambda_\textrm{D} &=0.00307\bigl(\zeta_\textrm{LB} - 0.0629\bigr).
    \label{eqn:D_mode_eigenvalue_numerical_units}
\end{align}
To acquire the growth rates numerically, we individually seeded S- and D-modes with the forms given by~\eqref{eqn:theta_S_linear} and~\eqref{eqn:theta_D_linear} respectively, and then analysed their evolution at different values of $\zeta_\textrm{LB}$. To extract their linear growth rates, we plotted the maximum mode amplitude over time on a logarithmic scale and used the gradient to extract the mode's exponential growth rate. We plot the exponential growth rates vs activity in Fig~\ref{fig:growth-rates-single}, from which we can extract the growth rate coefficient and the activity threshold. 

We see a very good agreement with both the growth rate magnitude and the threshold for both the S- and D-mode, with percentage differences not exceeding 5\%. One notable discrepancy is the threshold prediction from the graph of the D-mode's growth rate, which is larger than that of the S-mode yet is predicted to be the same from the linear stability analysis. Around the threshold, we observe a plateau before it eventually starts to grow linearly, the fit of which gives us the threshold prediction. The simulations for a seeded D-mode were significantly more difficult and unstable compared to the S-mode, especially around threshold, making D-mode's threshold result unsurprising. This same anomalous behaviour for a seeded D-mode was also observed by Marenduzzo \textit{et al.}~\cite{marenduzzo2007}.  As expected, the growth rates eventually deviate from a linear relation at a large enough activity above threshold.

The difference in the numerical values of the two growth rates suggests a separation of timescales that allows us to treat the instability as effectively a two-stage process. Initially, the S-mode grows fastest and attains a finite amplitude and steady state, while the D-mode remains infinitesimal. Subsequently, the D-mode evolves on top of the established S-mode. As the S-mode spontaneously breaks rotational symmetry within the cell, the problem is no longer isotropic and we consider separately growth of the nascent D-mode parallel and perpendicular to the established S-mode. 

\begin{figure}[tb]
 \centering
 \includegraphics[width=\columnwidth]{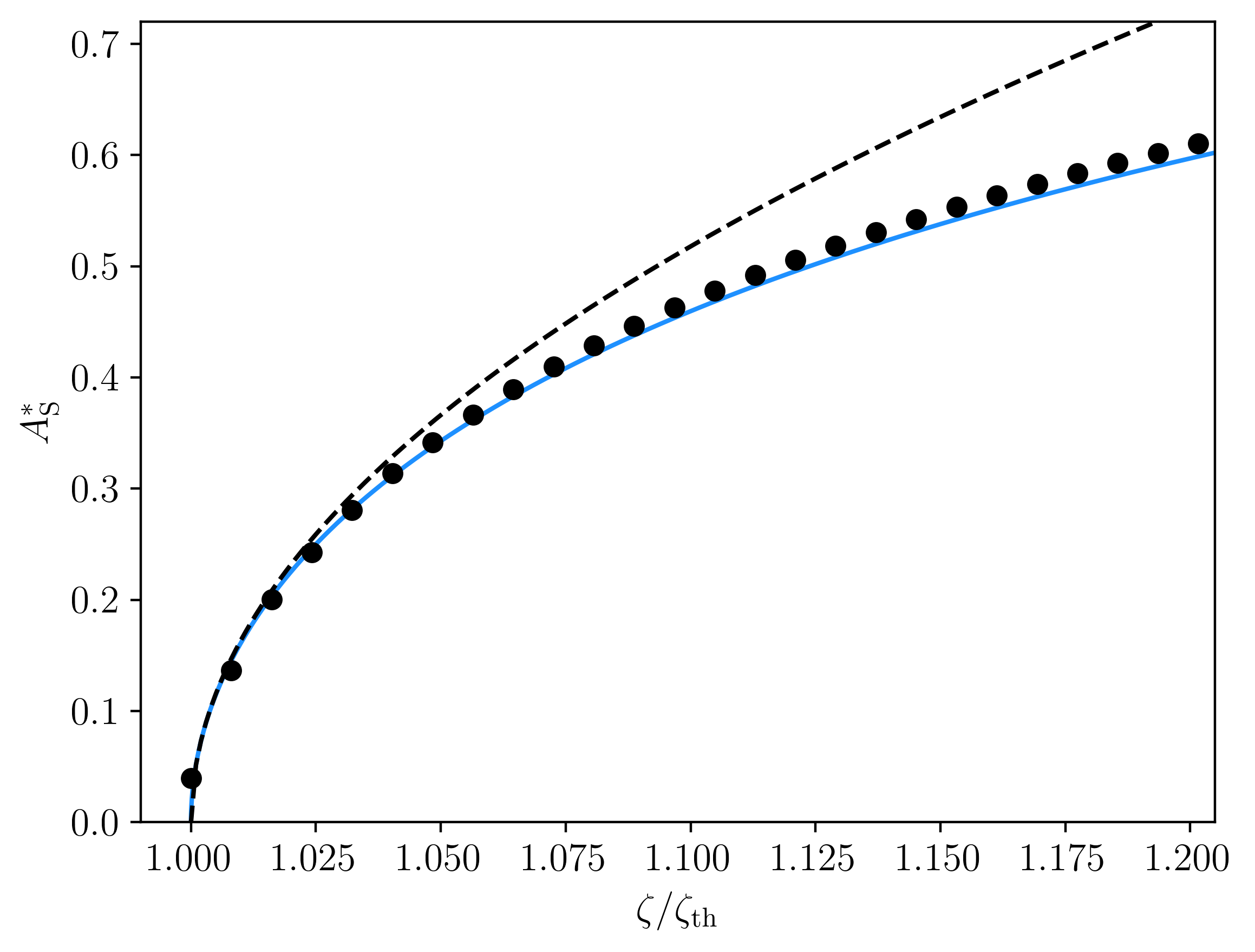}
 \caption{Plots of the S-mode amplitude against $\zeta/\zeta_\textrm{th}$. The blue line shows the quadrature solution~\eqref{eqn:theta_quadrature} and the black, dashed line the leading order part of the expansion of $A_\textrm{S}^*$, given by~\eqref{eqn:S_mode_amp_from_integral}. The black data points show numerical data of an individual S-mode in steady state, using the reduced simulation box of 3 bulk points perpendicular to the direction of flow. The numerical data is scaled with a threshold activity of $\zeta_\textrm{th}=0.062$ to fit the analytical prediction, again in good agreement with the theoretical prediction.}
 \label{fig:S-mode_amp}
 \end{figure}

We denote by $\theta^*(z)$ the steady state solution of~\eqref{eqn:theta_S_time_vector_form} with $\varphi = 0$, corresponding to a fully established pure S-mode. It is given by 
\begin{equation}
    \frac{K}{\gamma} \partial_z^2 \theta^* + \frac{\zeta\bigl(1 - \nu \cos 2\theta^*\bigr) \sin 2\theta^*}{4\mu+\gamma\bigl(1-\nu\cos2\theta^*\bigr)^2} = 0 ,
\label{eqn:full_non_linear_single_theta}
\end{equation}
and reduces to the quadrature 
\begin{equation}
    \frac{|z-d/4|}{\sqrt{2|\nu|K/\zeta}} = \int_{\theta^{*}}^{A_\textrm{S}^*} \biggl[ \ln \frac{1 + \frac{\gamma}{4\mu} (1 - \nu \cos 2\theta^{\prime})^2}{1 + \frac{\gamma}{4\mu} (1 - \nu \cos 2A_{\textrm{S}}^*)^2} \biggr]^{-1/2} \,d\theta^{\prime} ,
    \label{eqn:theta_quadrature}
\end{equation}
where the expression applies for $0 \leq z \leq d/2$; for $d/2 \leq z \leq d$ we use the odd symmetry $\theta^{*}(z) = - \theta^{*}(d-z)$. The amplitude of the mode is $A_{\textrm{S}}^* = \theta^{*}(d/4)$, which may be obtained implicitly as a function of $\zeta$ by setting $z = \theta^{*} = 0$ in~\eqref{eqn:theta_quadrature}. We show this dependence in Fig.~\ref{fig:S-mode_amp}. The behaviour close to threshold has the square root form $A_{\textrm{S}}^* \sim (\zeta - \zeta_{\textrm{th}})^{1/2}$, which may be found from an expansion of~\eqref{eqn:theta_quadrature} (with  $z = \theta^{*} = 0$) to linear order in $A_{\textrm{S}}$. Explicitly, to leading order we find the amplitude is  
\begin{equation}
    {A_\textrm{S}^*}^{2} = \frac{2(1-\nu)\Bigl(1+\frac{\gamma(1-\nu)^2}{4\mu}\Bigr)}{1-4\nu+\frac{\gamma(1-\nu)^2(1+2\nu)}{4\mu}} \frac{(\zeta-\zeta_\textrm{th})}{\zeta_\textrm{th}} . 
    \label{eqn:S_mode_amp_from_integral}
\end{equation}

We now determine the growth rate of the D-mode, to linear order in $\zeta - \zeta_{\textrm{th}}$, in the presence of a steady state S-mode. This amounts to retaining all terms up to $O({A_\textrm{S}^*}^{2})$ from the steady-state S-mode in the linearised dynamics for the D-mode, which therefore modifies the growth rates as compared to~\eqref{eqn:D_mode_eigenvalue}. We consider separately the growth of the D-mode parallel and perpendicular to the (spontaneously chosen) direction of the established S-mode. For the perpendicular case we substitute $\theta = \theta^*(z)$, $\varphi = \delta\varphi_{\textrm{D}}(z,t)$ into~\eqref{eqn:theta_D_time_vector_form} and linearise in $\delta\varphi_{\textrm{D}}$. The calculation of the growth rate uses the same perturbation theory as before and is given in Appendix~\ref{app:anisotropic}. For the parallel case we substitute $\theta = \theta^*(z) + \delta\theta_{\textrm{D}}(z,t)$ into~\eqref{eqn:theta_S_time_vector_form} and linearise in $\delta\theta_{\textrm{D}}$; the analysis is again given in Appendix~\ref{app:anisotropic}. The two growth rates are 
\begin{align}
\lambda_{\perp} & = \frac{4-2\nu+\frac{\gamma}{2\mu}(1-\nu)^2(2+\nu)}{1-4\nu+\frac{\gamma}{4\mu}(1-\nu)^2(1+2\nu)} \,\lambda_\textrm{D} , \label{eqn:perp_growth_rate} \\
\lambda_{\parallel} & = \frac{8\nu+\frac{2\gamma}{\mu} \nu(1-\nu)^2}{1-4\nu+\frac{\gamma}{4\mu}(1-\nu)^2(1+2\nu)} \,\lambda_\textrm{D} , \label{eqn:parr_growth_rate}
\end{align}
which in simulation units read  
\begin{align}
\lambda_{\perp} & = 0.00747 \bigl( \zeta_\textrm{LB} - 0.0629 \bigr) , \\ %\label{eqn:theta_D_mode_anisotropic_growth}\\
\lambda_{\parallel} & = -0.0166 \bigl( \zeta_\textrm{LB} - 0.0629 \bigr) . %\label{eqn:theta_S_mode_anisotropic_growth}
\end{align}
The main result is that $\lambda_\parallel<0$ and $\lambda_\perp>0$ so that the D-mode only remains linearly unstable along the direction perpendicular to that set by the S-mode. As a result, the director evolves into a truly three-dimensional configuration with the S- and D-modes growing along orthogonal in-plane directions. This interplay between the two modes leads to twisted director fields with 
\begin{align}            
    \vb{n} \vdot \curl \vb{n} & = \cos\theta \cos\varphi \sin\varphi \,\partial_z \theta - \sin\theta \,\partial_z\varphi , \\
    & \approx - \frac{\pi}{d} \,A_{\textrm{S}} A_{\textrm{D}} \biggl( 1 - \cos\frac{2\pi z}{d} \biggr) ,
\end{align}
where in the second form we have linearised in $\theta$ and $\varphi$ and taken them to have the threshold forms~\eqref{eqn:theta_S_linear} and~\eqref{eqn:theta_D_linear}, respectively. The twist maintains a single sign (handedness) throughout the cell, vanishing only on the two boundaries. Since the S-mode spontaneously breaks rotational symmetry in the plane, its amplitude $A_{\textrm{S}}$ is always positive. In contrast, the amplitude of the D-mode, $A_{\textrm{D}}$, can be positive or negative (corresponding to the two directions orthogonal to the established S-mode); when it is positive the twist is right-handed and when negative it is left-handed. In a nematic material we expect both to occur with equal probability and any particular realisation represents a spontaneous chiral symmetry breaking. This general mechanism for confined active nematics may also be relevant to the emergence of twist in bulk three-dimensional systems~\cite{shendruk2018twist,Chandrakar2020bend} and possibly also to the prevalence of twist loops in the statistics of their defect loops \cite{duclos2020topological,houston2022defect}. 

\begin{figure}[tb]
 \centering
 \includegraphics[width=\columnwidth]{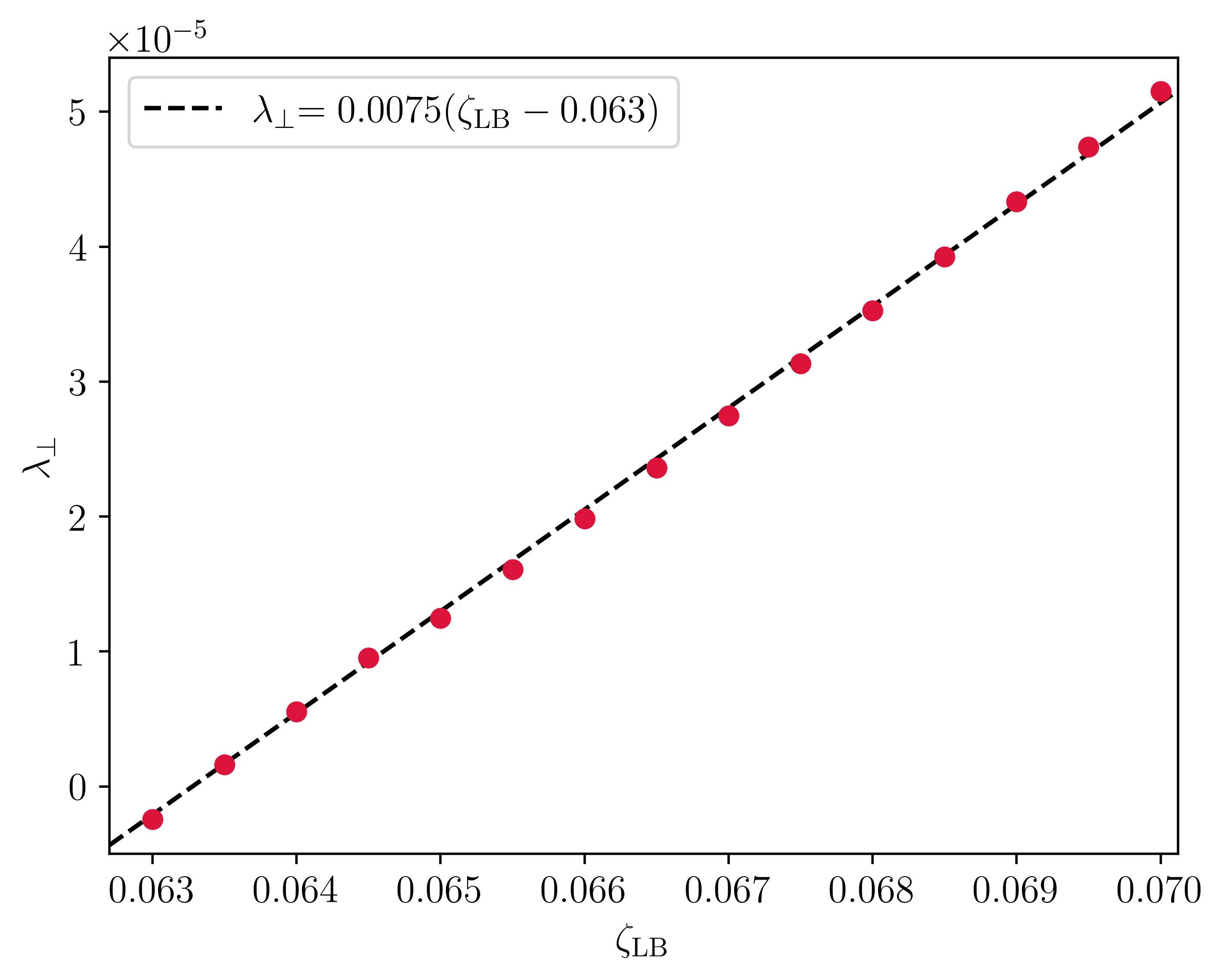}
 \caption{Plot of the D-mode's growth rate perpendicular to an established S-mode. The black dashed line is a linear fit of the numerical data in the selected range of data close to the threshold.}
 \label{fig:anisotropic_perturbations_growth}
 \end{figure}
 \begin{figure*}[t]
    \centering
    \includegraphics[width=\linewidth]{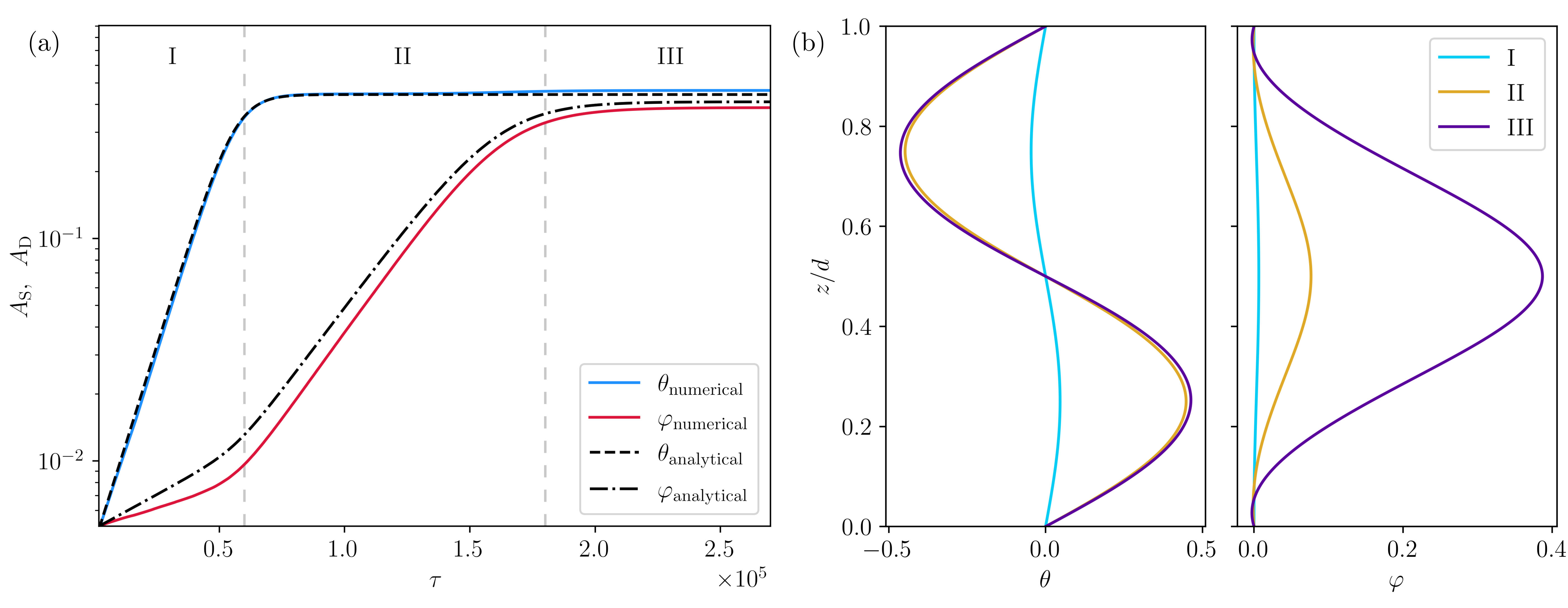}
    \caption{(a) Comparison of the numerical evolution of the S-mode amplitude and D-mode amplitude (blue and red lines, respectively), with the analytical predictions of the leading order amplitude evolution from~\eqref{eqn:A_S_evolution} and~\eqref{eqn:A_D_evolution} (black, dashed lines) at $\zeta_\textrm{LB}=0.0675$. The plot has a logarithmic scale on the vertical axis. (b) Left: S-mode profile along the cell gap at three different times of the time evolution (I, II and III, as seen in (a)). Right: D-mode profile along the cell gap at the same three time points. We used time frames $0.3 \times 10^5 \tau$, $1.2 \times 10^5 \tau$ and $2.7 \times 10^5\tau$ for I, II and III respectively. 
    }
\label{fig:time evolution}
\end{figure*}

The growth rate $\lambda_{\perp}$~\eqref{eqn:perp_growth_rate} for the orthogonal D-mode can be verified numerically by initialising the director with an S-mode along ${\bf e}_x$ and a small amplitude D-mode along ${\bf e}_y$. Tracking the exponential growth of the D-mode as a function of activity allows for a fit of the growth rate and threshold activity as before. This is shown in Fig.~\ref{fig:anisotropic_perturbations_growth}. The agreement with the theoretical prediction is again excellent. We note, particularly, that we obtain better agreement for the threshold activity $\zeta_{\textrm{th}}$ than we found from simulations with only the D-mode.

\section{Mode Evolution and Steady State}
\label{sec:analytical_amplitude_evolution}

We now summarise and describe the full evolution of the instability to the steady spontaneous flow state. This can be studied systematically in numerics by seeding a small amplitude director perturbation consisting of an S-mode along ${\bf e}_x$ and a D-mode along ${\bf e}_y$ and tracking their amplitudes -- the maximum values of $\theta$ and $\varphi$ -- over time. This is shown in Fig.~\ref{fig:time evolution}. The evolution can be divided into three distinct regimes: in the first (I), there is exponential growth of both modes, but with the S-mode growing significantly faster. This corresponds to the independent and isotropic mode dynamics described in Fig.~\ref{fig:growth-rates-single}. In the second regime (II), the S-mode amplitude attains a plateau and there is an increase in the exponential growth rate of the D-mode. The S-mode amplitude at its plateau corresponds to the value $A_{\textrm{S}}^{*}$ described in \S\ref{sec:evol_above_thresh} and the enhanced growth rate of the D-mode is the cross-over to the rate $\lambda_{\perp}$ as described in Fig.~\ref{fig:anisotropic_perturbations_growth}. Finally, in the third regime (III) the D-mode amplitude attains its steady state value and promotes a small further increase of the S-mode amplitude to its steady state. 

\begin{figure}[t]
    \centering
    \includegraphics[width=\columnwidth]{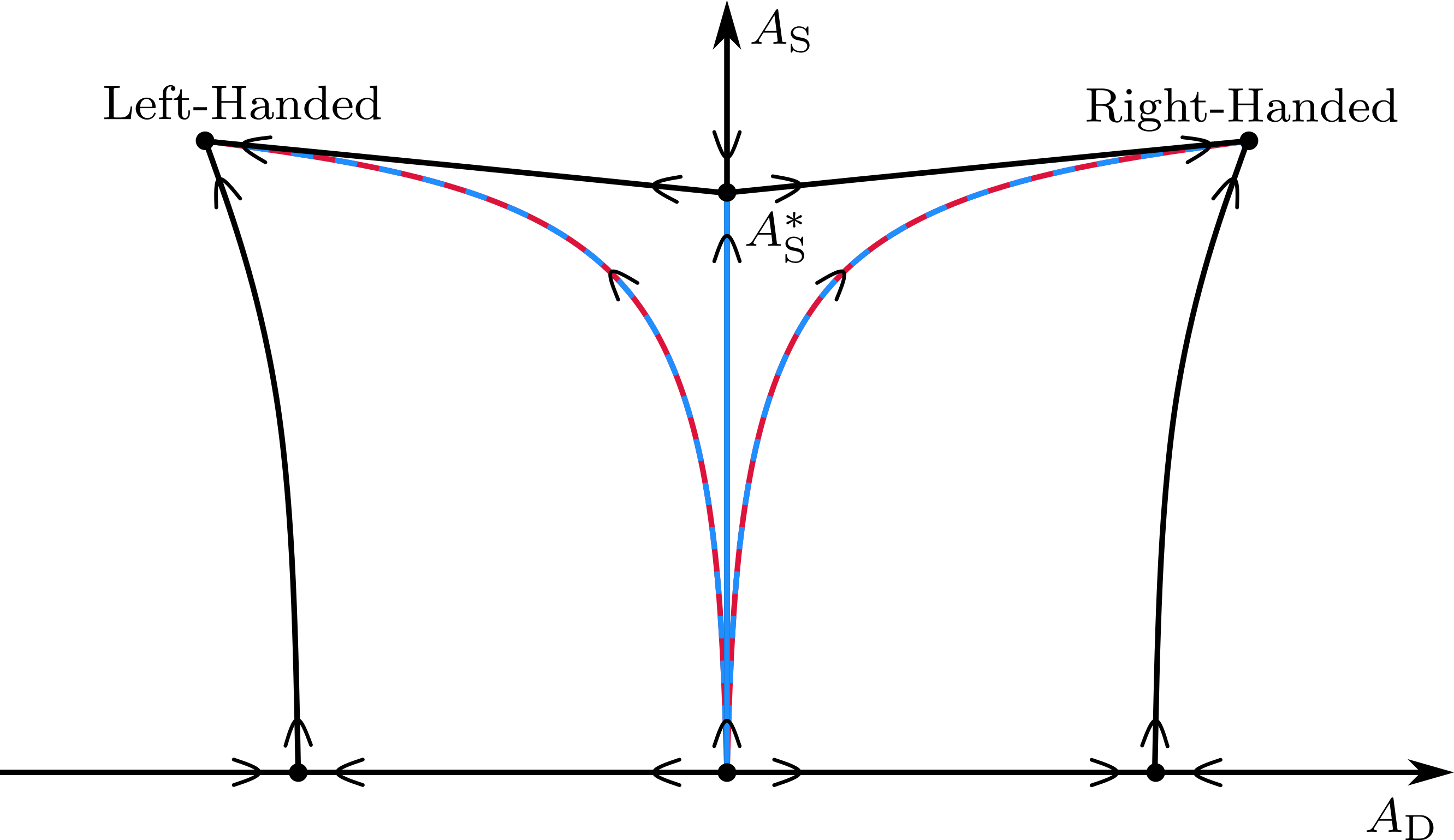}
    \caption{Schematic of the phase portrait of the spontaneous flow transition above threshold. The vertical blue line is the S-mode trajectory where the S-mode grows independently of the D-mode, which is the approximation made in~\eqref{eqn:decoupling_approximation}. The red and blue dotted lines show the trajectory of the coupled evolution of the S- and D-modes to the left- and right-handed chiral states.}
\label{fig:phase prtrait}
\end{figure}

This joint evolution can be cast as a coupled dynamical system for the amplitudes $A_{\textrm{S}}, A_{\textrm{D}}$ of the S- and D-modes 
\begin{align}
    & \frac{dA_{\textrm{S}}}{dt} = g_{\textrm{S}}\bigl( A_{\textrm{S}} , A_{\textrm{D}} \bigr) , 
    && \frac{dA_{\textrm{D}}}{dt} = g_{\textrm{D}}\bigl( A_{\textrm{S}} , A_{\textrm{D}} \bigr) , 
\end{align}
where the growth rate functions $g_{\textrm{S}} , g_{\textrm{D}}$ have the fixed point structure shown schematically in Fig.~\ref{fig:phase prtrait}. We define $A_\textrm{S}$ to be strictly positive, meaning that $A_\textrm{D}$ can take either sign. There are three important fixed points: the origin and the two points corresponding to the right- and left-handed states. Below threshold, the origin is a stable fixed point, but above it becomes unstable to all S- and D-mode perturbations. Depending on the sign of $A_\textrm{D}$, perturbations around the origin will either flow to the left-handed or the right-handed stable fixed points, corresponding to the handedness of the resulting flow state's chirality. These flows are shown by the blue and red dashed lines in Fig.~\ref{fig:phase prtrait}. The trajectory follows closely to the $A_\textrm{S}$ axis until $A_\textrm{S}$ is large and then rapidly moves away from the axis to the fixed points. This corresponds physically to the S-mode  growing to a large amplitude before there is any significant D-mode growth. Finally, we note that in the absence of a D-mode, the S-mode grows to the semi-stable fixed point labelled as $A_\textrm{S}^*$, which has a slightly smaller amplitude than the left- and right-handed fixed points. This fixed point is described in~\ref{eqn:theta_quadrature}. For activities close to threshold, the fixed points are close to the origin and we can expand the growth functions as
\begin{align}
    g_{\textrm{S}}\bigl( A_{\textrm{S}} , A_{\textrm{D}} \bigr) & = \lambda_{\textrm{S}} A_{\textrm{S}} - \Lambda_1 A_{\textrm{S}}^3 + \Lambda_2 A_{\textrm{S}} A_{\textrm{D}}^2 + \cdots , \label{eqn:growth_function_S} \\
    g_{\textrm{D}}\bigl( A_{\textrm{S}} , A_{\textrm{D}} \bigr) & = \lambda_{\textrm{D}} A_{\textrm{D}} + \Lambda_3 A_{\textrm{S}}^2 A_{\textrm{D}} - \Lambda_4 A_{\textrm{D}}^3 + \cdots ,
    \label{eqn:growth_function_D}
\end{align}
where the non-linear terms are those allowed by symmetry. We give the calculation of the $\Lambda$ coefficients in Appendix~\ref{app:steady_state}. This system connects to our previous results and reproduces the numerically observed dynamics of Fig.~\ref{fig:time evolution}. The amplitude $A_{\textrm{S}}^{*}$ of the S-mode plateau in regime II is given by $(\lambda_{\textrm{S}}/\Lambda_1)^{1/2}$ and matches the value in~\eqref{eqn:S_mode_amp_from_integral}. Similarly, the enhanced growth rate of the D-mode in regime II is given by $\lambda_{\textrm{D}} + (\Lambda_3/\Lambda_1) \lambda_{\textrm{S}}$ and matches the rate $\lambda_{\perp}$ in~\eqref{eqn:perp_growth_rate}. At this leading order, we obtain the steady state amplitudes
\begin{align}                   
    A_\textrm{S} = & \sqrt{\frac{\Lambda_4\lambda_\textrm{S} + \Lambda_2\lambda_\textrm{D}}{\Lambda_1\Lambda_4 - \Lambda_2\Lambda_3}} \propto \sqrt{\zeta-\zeta_\textrm{th}},\\    A_\textrm{D} = & \sqrt{\frac{\Lambda_1\lambda_\textrm{D} + \Lambda_3\lambda_\textrm{S}}{\Lambda_1\Lambda_4 - \Lambda_2\Lambda_3}} \propto \sqrt{\zeta-\zeta_\textrm{th}}.
\end{align}
We note that 
\begin{equation}
    A_\textrm{S}^2 = {A_\textrm{S}^*}^2 + \frac{\Lambda_2}{\Lambda_1} A_\textrm{D}^2 .
\end{equation}
This coincides with the numerical observation that there is an increase in the S-mode amplitude when the D-mode comes into steady state. In the numerical observations, this increase in small, implying that the D-mode coupling to the S-mode evolution is small, i.e. $\Lambda_2/\Lambda_1\ll 1$. Indeed, in simulation units, $\Lambda_2/\Lambda_1=0.08$. With this weak coupling, the evolution of the S-mode can be approximated by
\begin{equation}
    \dv{A_\textrm{S}}{t}\approx\bigl(\lambda_\textrm{S}-\Lambda_1A_\textrm{S}^2\bigr)A_\textrm{S},\label{eqn:decoupling_approximation}
\end{equation}
which gives
\begin{equation}
    A_\textrm{S}(t) = A_\textrm{S}(0) \,e^{\lambda_\textrm{S} t} \Biggl[ 1 + \frac{ A_\textrm{S}(0)^2}{{A_\textrm{S}^*}^2} ( e^{2\lambda_\textrm{S} t} - 1 ) \Biggr]^{-1/2} .
    \label{eqn:A_S_evolution}
\end{equation}
We can substitute this into the leading order part of $g_\textrm{D}$ and solve to give the approximate the evolution of $A_\textrm{D}$ as
\begin{equation}
    \begin{split}
        A_{\textrm{D}}(t) & = A_{\textrm{D}}(0) \,e^{\lambda_{\textrm{D}} t} \Biggl[ 1 + \frac{ A_\textrm{S}(0)^2}{{A_\textrm{S}^*}^2}\bigl( e^{2\lambda_{\textrm{S}} t} - 1 \bigr) \Biggr]^{\Lambda_3/2\Lambda_1} \\
        & \times \Biggl[ 1 + \frac{A_{\textrm{D}}(0)^2 \Lambda_4}{\lambda_{\textrm{D}}} \biggl( 1 - \frac{ A_\textrm{S}(0)^2}{{A_\textrm{S}^*}^2} \biggr)^{\Lambda_3/\Lambda_1} \\
        & \biggl( e^{2\lambda_{\textrm{D}} t} \mathbin{_2F_1}\!\biggl[\frac{\lambda_\textrm{D}}{\lambda_\textrm{S}},-\frac{\Lambda_3}{\Lambda_1};1+\frac{\lambda_\textrm{D}}{\lambda_\textrm{S}};\frac{A_\textrm{S}(0)^2e^{2\lambda_\textrm{S} t}}{A_\textrm{S}(0)^2-{A_\textrm{S}^*}^2}\biggr] \\
        & - \mathbin{_2F_1}\!\biggl[\frac{\lambda_\textrm{D}}{\lambda_\textrm{S}},-\frac{\Lambda_3}{\Lambda_1};1+\frac{\lambda_\textrm{D}}{\lambda_\textrm{S}};\frac{A_\textrm{S}(0)^2}{A_\textrm{S}(0)^2-{A_\textrm{S}^*}^2}\biggr] \biggr) \Biggr]^{-1/2} ,
    \end{split}
    \label{eqn:A_D_evolution} 
\end{equation}
where $_2F_1$ is Gauss's hypergeometric function.
 
We compare \eqref{eqn:A_S_evolution} and \eqref{eqn:A_D_evolution} with numerical data of mode amplitude evolution in Fig. \ref{fig:time evolution} for $\zeta=0.0675$. The analytical model captures the qualitative, triphasic nature of the system well. For the S-mode, the agreement is very good up to phase III, where the second plateau is not captured due to the decoupling approximation that was made. For the D-mode, the analytical prediction of the growth rate in phase I, \eqref{eqn:D_mode_eigenvalue_numerical_units}, is  larger than the numerics, although this is consistent with what we have already observed when we seeded an independent D-mode. Recall that the numerical isotropic growth rate for a D-mode has a shifted threshold compared to \eqref{eqn:D_mode_eigenvalue_numerical_units} which has a significant effect on the growth rate close to threshold, thus making the numerical growth rate noticeably smaller than what is analytically predicted. Nevertheless, the analytical prediction tracks the numerical data very well thereafter, albeit translated upwards due to the first phase growth rate being too large.

As we move further and further above threshold, the analytical model becomes worse and worse. If, however, $\theta$ and $\varphi$ are in odd end even symmetry classes respectively, then the functions $\partial_t\theta$ and $\partial_t\varphi$ are also odd and even respectively which can be checked by inspection of each term in~\eqref{eqn:theta_S_time_vector_form} and~\eqref{eqn:theta_D_time_vector_form}. This implies that if $\theta$ and $\varphi$ start off as odd and even respectively, then the functions will remain in the same symmetry class for the entirety of their non-linear, coupled time evolution. Furthermore, we can analyse the symmetry of $\vb{v}$ and upon inspection of the formula, if $\theta$ and $\varphi$ are in their aforementioned symmetry classes then $v_x$ will be an even function and $v_y$ will be and odd function. Hence, we expect that the generation of a chiral director field is a general property of the system, rather than just a feature close to threshold.

\section{Discussion and conclusions}
\label{sec:conclusion}

We have studied a spontaneous flow transition in an active nematic fluid with an infinite slab geometry and normal surface anchoring. We find the existence of two independent flow instabilities, the S-mode and the D-mode, that occur at the same threshold but have different growth rates above threshold, which we calculate using perturbation theory of a non-Hermitian integro-differential operator. Above threshold, the S-mode with its larger growth rate grows in a random direction to steady state, breaking the initial rotational symmetry of the system. Thereafter, any perturbations within the system are subject to an anisotropic environment. In particular, D-mode perturbations that are parallel and perpendicular to the direction of anisotropy decay and grow respectively. The growth of the D-mode perpendicular to the S-mode yields a steady-state with a full, three-dimensional chiral director field, with spontaneously broken chiral symmetry. We analytically describe the mode growth with a leading-order model that captures the key characteristics of the system and its evolution into the spontaneously flowing state.

The first natural extension of this work is to explore the possible inhomogeneity of the flow within the cell plane, enabling the study of the chiral flowing state's stability to the Goldstone mode coming from the breaking of rotational symmetry and the umbilic defect lines associated to this. Furthermore, in large systems, the spontaneous chiral symmetry breaking could yield domains of different chirality, leading to an effective non-conserved binary mixture. We hope the novel three-dimensional flow instability we have uncovered can provide motivation for experimental research of active nematic systems with normal anchoring. 

Active nematics are fundamentally analogous to passive (i.e. non-active) nematic liquid crystals, with the orientational ordering of the anisotropic material building blocks crucially determining the material dynamics, including at the surface. Today, surface anchoring in passive nematics can be realised experimentally in different configurations, ranging from uniform planar and degenerate planar to homeotropic and even tilted and tilted degenerate. Such advanced control over surface alignment was shown together with control over confinement geometries to diverse material structures and phenomena, such as realisation of colloidal and field knots \cite{tkalec2011knots,Martinez2014knots, MachonAlexander2014knots}, self-assembly \cite{self-assembly}, hexadecapolar and 32-pole field configurations \cite{hexadecapolar, multipoles}, memory \cite{memory-araki,memory-serra}, static and dynamic solitons \cite{solitons, solitons2, solitons3}, tunable positioning of topological defects \cite{tunabledefects,holestericstripes}. Clearly, advancing the ability to control different surface anchoring regimes in combination with confinement \cite{confinement-organization} could open diverse research directions in confined active nematics, especially at the experimental level. Even rather simple surfaces like spheres imposing homeotropic anchoring on active nematics would lead to the emergence of topologically imposed and conditioned bulk Saturn ring defect states that are commonly observed in passive nematics but not seen in active nematics. For example, could combining three-dimensional self-assembly of (active) nano-objects (e.g., bacteria) combined with confinement lead to different effective anchoring beyond the typical planar? \cite{confinement} 

Another possibly emerging direction from this work is the observation of the spontanenous chiral symmetry in an active (nematic) system. At least in non-active nematic materials, chiral symmetry breaking naturally leads to the emergence of different phenomena, such as chiral domains and associated topological defects such as solitonic-like nematicons \cite{symmetry-breaking}. In active systems, chiral activity can lead to topological edge modes \cite{souslov2017} and odd elastic responses \cite{scheibner2020}. 
Combining spontaneous chiral symmetry breaking with activity in more advanced geometries and setups -- beyond the simple homeotropic cells studied in this work -- could lead to an exciting advancement of the control and design of novel active functional matter.

\section{Methods}
\label{sec:methods}

\subsection{Numerical Methods}

We solve the Beris-Edwards equations~\eqref{eqn:continuityN}-\eqref{eqn:Beris-Edwards} using a hybrid lattice Boltzmann algorithm. In these equations, $\vb*{\Pi}$ is the stress tensor defined by
\begin{equation}
\begin{split}
    \Pi_{ij}=&-p\delta_{ij}+\mu(\partial_iv_j+\partial_j v_i)+ Q_{ik} H_{jk}-H_{ik} Q_{jk} \\
    &+ 2 \chi \left( Q_{ij}+\frac{1}{3} \delta_{ij} \right) Q_{kl} H_{kl}-\chi H_{ik} \left( Q_{kj} +\frac{1}{3} \delta_{kj}\right)\\
    &-\chi \left( Q_{ik} +\frac{1}{3} \delta_{ik} \right) H_{kj}-\partial_i Q_{k l} \frac{\delta F}{\delta \partial_j Q_{k l}}-\zeta_\textrm{LB} Q_{ij} ,
\end{split}
\end{equation}
where $p$ is the pressure, $\mu$ is an isotropic shear viscosity, $\chi$ is the flow alignment parameter, $\Gamma$ is a rotational viscosity and $\zeta_{\textrm{LB}}$ is the activity parameter. $H_{ij}=-\frac{\delta F}{\delta Q_{ij}}+\frac{1}{3} \delta_{ij} \mathrm{Tr} \left( \frac{\delta F}{\delta Q_{kl}}\right)$ are the components of the molecular field, in which
\begin{equation}
\begin{split}
    F = & \int \mathcal{F}_\textrm{B}\, \dd{V} + \int \mathcal{F}_\textrm{S}\, \dd{S},
\end{split}
\end{equation}
is the free energy, where
\begin{align}
    \mathcal{F}_\textrm{B}\!=&\frac{A}{2} Q_{ij} Q_{ji}\!+\!\frac{B}{3} Q_{ij} Q_{jk} Q_{ki}\!+\!\frac{C}{4} (Q_{ij}Q_{ij})^2\!+\!\frac{L}{2}\! \left(\partial_k Q_{ij} \right)^2\!\!\!, \\
    \mathcal{F}_\textrm{S}=&\frac{1}{2} W_{\mathrm{h}} \left(Q_{ij}-Q^0_{ij}\right)^2.
\end{align} 
The bulk part of the free energy is described via phenomenological constants for phase transition $A$, $B$, $C$, one constant approximation for the elastic part $L$, and the surface part is described via homeotropic anchoring with strength $W_\mathrm{h}$ and orientation $Q^0_{ij}$ corresponding to the preferred perpendicular director on the surface plate $\vb{n}=\pm\vb{e}_z$. Finally, the tensor $\vb{S}$ with components 
\begin{equation}
\begin{split}
     S_{ij}=&\left(\chi {D_{ik}}-{\Omega_{ik}}\right)\left(Q_{kj}+\frac{1}{3}\delta_{kj}\right)\\
      &+\left(Q_{ik }+\frac{1}{3}\delta_{ik}\right)\left(\chi {D_{kj}} + {\Omega_{kj}}\right)\\
     &-2\chi \left(Q_{ij }+\frac{1}{3}\delta_{ij}\right) Q_{kl} D_{lk},
\end{split}
\end{equation}
describes the coupling between the nematic order parameter and the flow field. We use $\vb{D}$ and $\vb*{\Omega}$ to represent the symmetric and antisymmetric parts of the velocity gradient tensor respectively.

In the lattice Boltzmann simulations, the fundamental scaling parameters are nematic correlation length 
\begin{equation}
    \xi_n= \sqrt{L/({A+ \frac{9}{2} B s_{eq}+C s^2_{eq}})}
\end{equation}
and nematic time scale
\begin{equation}
    \tau_n = \xi^2_n/\Gamma L.
\end{equation}
All of the parameter values are expressed in units of the elastic constant $L$. We used the following Landau-de Gennes parameters $A=-0.687\, \, \, L/\xi_n^2$, $B=-3.53\, \, \, L/\xi_n^2$ and $C=2.89\, \, \, L/\xi_n^2$, describing the phase transition part of the free energy and the rotational viscosity $\Gamma=\frac{\xi^2_n}{\tau_n L}$. Under the aforementioned choice of parameters $s_{eq}=0.651$. This corresponds to the use of Beris-Edwards parameters $\chi=1$, $\mu=1.38/\Gamma$ and $\rho=0.031 \,\,\, \frac{1}{L\Gamma^2}$. We performed simulations with a cell size of $201\times 201\times 45$ bulk points confined between the two plates using homeotropic anchoring with strength $W_{h}=\frac{2}{3} \,  L/\xi_n$ and periodic boundary conditions on the side. We used disretisation of spatial coordinates $\Delta x= 1.5\, \xi_n$ and a time step $\Delta t= 0.025 \, \tau_n$.

The nematic director, $\vb{n}$, is obtained as the eigenvector associated with the largest eigenvalue of $\vb{Q}$, which represents the magnitude of the order parameter, $S$.

\subsection{Analytical Methods}

\subsubsection{Details of the Ericksen-Leslie Equations}

In the  Ericksen-Leslie equations, \eqref{eqn:continuity}, \eqref{eqn:Stokes_n} and \eqref{eqn:n_time_evolution},
$\gamma =\frac{9S^2}{2\Gamma}$ and $\nu = -\frac{(3S+4)}{9S}\chi$. $h_i=-\fdv{F}{n_i}+\fdv{F}{n_j}n_jn_i$ is the molecular field, in which we use the Frank free energy with the one-elastic constant approximation, 
\begin{equation}
    F=\frac{K}{2}\int\partial_in_j\partial_in_j\,\dd V,
\end{equation}
where $K=\frac{9S^2}{2}L$. We explicitly separate out the stress tensor. The elastic contribution is 
\begin{equation}
    \sigma^{\textrm{el}}_{ij}= \frac{1}{2} \bigl( n_i h_j - h_i n_j \bigr)+\frac{\nu}{2} \bigl( n_i h_j + h_i n_j \bigr)  -K\partial_in_k\partial_j n_k
\end{equation}
and the active contribution, $\sigma^{\textrm{a}}_{ij}=-\zeta n_in_j$, where $\zeta=\frac{3S}{2}\zeta_\textrm{LB}$. 

\subsubsection{Eigenvalues above Threshold}
\label{sec:eigenvalues_above_threshold}

We solve for the leading order correction to the eigenvalue, $\lambda$, of the stability operator equation $\mathcal{L}\theta=\lambda\theta$ when $\zeta>\zeta_\textrm{th}$. To do so, we perform a perturbation expansion, $\mathcal{L}=\mathcal{L}_0+\mathcal{L}_1+\dots$ (and likewise for $\lambda$ and $\theta$), in powers of $\zeta-\zeta_{\textrm{th}}$, with $\mathcal{L}_0$ corresponding to $\mathcal{L}$ when $\zeta=\zeta_\textrm{th}$ (and likewise for $\lambda_0$ and $\theta_0$). In the main text, $\mathcal{L}$ is defined by \eqref{eqn:linear_theta_operator}. However, it is important to note that the analysis provided in this section is valid for all operators, $\mathcal{L}$, that linearise to 
\begin{equation}
\begin{split}
     \mathcal{L}_0\theta=\frac{K}{\gamma} \partial_{z}^2 \theta + \frac{K(1-\nu)^2}{4\mu} \Bigl( \partial_{z}^2 \theta &-\bigl\langle \partial_{z}^2 \theta \bigr\rangle \Bigr)\\
     &+ \frac{\zeta_\textrm{th}(1-\nu)}{2\mu} \Bigl( \theta - \langle\theta \rangle \Bigr),
    \label{eqn:threshold_operator}
\end{split}
\end{equation}
when $\zeta=\zeta_\textrm{th}$. We proceed in the usual way of expanding $\mathcal{L}\theta=\lambda\theta$ and equating order by order, giving
\begin{gather}
    \mathcal{L}_0\theta_0=\lambda_0\theta_0=0\\
    \mathcal{L}_0\theta_1+\mathcal{L}_1 \theta_0 = \lambda_0\theta_1+ \lambda_1\theta_0=\lambda_1\theta_0. \label{eqn:pertubation_theory_1}\\
    \vdots \nonumber
\end{gather}
$\mathcal{L}_0$ is non-Hermitian over the interval $z\in[0,d]$ due to the $\langle\partial_z^2\,\cdot\,\rangle$ term, meaning that we cannot apply standard methods for Hermitian operators to solve for $\lambda_1$; we must instead take a slightly different approach that is unique to operator \eqref{eqn:threshold_operator}. Firstly, we take the average of all terms in~\eqref{eqn:pertubation_theory_1} and rearrange to get
\begin{equation}
   \left\langle \mathcal{L}_0\theta_1\right\rangle= \frac{K}{\gamma}\left\langle \partial_z^2\theta_1\right\rangle= \lambda_1\left\langle\theta_0\right\rangle-\left\langle\mathcal{L}_1 \theta_0\right\rangle.\label{eqn:pertubation_theory_2}
\end{equation}
Next, we multiply \eqref{eqn:pertubation_theory_1} by $\theta_0$ and average, giving
\begin{equation}
\left\langle\theta_0\mathcal{L}_0\theta_1\right\rangle+\left\langle\theta_0\mathcal{L}_1 \theta_0\right\rangle= \lambda_1\left\langle\theta_0^2\right\rangle. \label{eqn:pertubation_theory_3}
\end{equation}
To simplify this, we make use the result
\begin{equation}
\begin{split}
    \left\langle\theta_0\mathcal{L}_0\theta_1\right\rangle \!=\!&\left\langle\theta_1\mathcal{L}_0\theta_0\right\rangle\!-\!\frac{K(1\!-\!\nu)^2}{4\mu}\!\Bigl(\!\left\langle\theta_0\right\rangle\!\left\langle\partial_z^2\theta_1\right\rangle\!-\!\left\langle\theta_1\right\rangle\!\left\langle\partial_z^2\theta_0\right\rangle\!\Bigr)\\=&-\frac{K(1-\nu)^2}{4\mu}\left\langle\theta_0\right\rangle\left\langle\partial_z^2\theta_1\right\rangle.
    \label{eqn:operator_commutator}
\end{split}
\end{equation}
where the second equality comes from the fact that $\left\langle\theta_1\mathcal{L}_0\theta_0\right\rangle=0$ and $\left\langle\partial_z^2\theta_0\right\rangle=0$. We can combine \eqref{eqn:pertubation_theory_2}, \eqref{eqn:pertubation_theory_3} and \eqref{eqn:operator_commutator} to eliminate $\left\langle\partial_z^2\theta_1\right\rangle$ and solve for $\lambda_1$, giving
\begin{equation}
    \lambda_1= \frac{\left\langle\theta_0\mathcal{L}_1 \theta_0\right\rangle+\frac{\gamma(1-\nu)^2}{4\mu}\left\langle\theta_0\right\rangle\left\langle\mathcal{L}_1 \theta_0\right\rangle}{\left\langle\theta_0^2\right\rangle+\frac{\gamma(1-\nu)^2}{4\mu}\left\langle\theta_0\right\rangle^2}.
    \label{eqn:eigenvalue_equation_full}
\end{equation}
We note that in the case where the mode is antisymmetric about the cell midplane, all of the integral terms in \eqref{eqn:threshold_operator} vanish, making $\mathcal{L}_0$ Hermitian over the interval $z\in[0,d]$. As expected, \eqref{eqn:eigenvalue_equation_full} reduces to $\lambda_1=\left\langle\theta_0\mathcal{L}_1 \theta_0\right\rangle/\left\langle\theta_0^2\right\rangle$, the standard result of perturbation theory on Hermitian operators. We use~\eqref{eqn:eigenvalue_equation_full} to calculate the isotropic and anisotropic growth rates, with calculation details given in Appendix~\ref{app:B}.

\subsubsection{Amplitude Evolution above Threshold}

We obtain the leading order time evolution equations of the S- and D-mode amplitudes close to threshold. We take $\theta$ to be an S-mode and $\varphi$ to be a D-mode, and then make the usual expansion $\theta=\theta_0+\theta_1+\dots$ and $\varphi=\varphi_0+\varphi_1+\dots$ in powers of $\zeta-\zeta_\textrm{th}$. We substitute these expansions into~\eqref{eqn:theta_S_time_vector_form} and~\eqref{eqn:theta_D_time_vector_form} and equate order by order. The leading order balance gives us the now familiar eigenfunctions which we write as $\theta_0=A_\textrm{S}(t)\sin\frac{2\pi z}{d}:=A_\textrm{S}(t)\psi_\textrm{S}(z)$ and $\varphi_0=\frac{A_\textrm{D}(t)}{2}\left(1-\cos\frac{2\pi z}{d}\right):=A_\textrm{D}(t)\psi_\textrm{D}(z)$. The subsequent analysis is given in terms of $\theta$, but is the same for $\varphi$. At the next order in $\zeta-\zeta_\textrm{th}$, we obtain
\begin{equation}
\psi\,\dv{A}{t}=\mathcal{L}_\textrm{1}\theta_0+\mathcal{L}_0\theta_\textrm{1},
\end{equation} 
We can now solve for $\dv{A}{t}$ in the same way as we solved for $\lambda_1$ in the previous sub-section, giving us
\begin{equation}
    \dv{A}{t}=\frac{\langle\psi\mathcal{L}_1 \theta_0\rangle+\frac{\gamma(1-\nu)^2}{4\mu}\langle\psi\rangle\langle\mathcal{L}_1 \theta_0\rangle}{\langle\psi^2\rangle+\frac{\gamma(1-\nu)^2}{4\mu}\langle\psi\rangle^2}.
    \label{eqn:amp_evol_general}
\end{equation}
This is the equation used to obtain the leading order parts of~\eqref{eqn:growth_function_S} and~\eqref{eqn:growth_function_D} in the main text. We go through the details of the calculations in Appendix~\ref{app:steady_state}.

\acknowledgements{For the purpose of open access, the authors have applied a Creative Commons Attribution (CC-BY) licence to any Author Accepted Manuscript version arising from this submission. We acknowledge support from Nordita during the Program on Current and Future Themes in Soft and Biological Active Matter. V.J.P. acknowledges funding from the EPSRC, grant EP/T51794X/1. M.R. and E.C. acknowledge funding from the Slovenian research agency ARRS grants P1-0099, N1-0195 and J1-2462, and EU ERC AdG LOGOS.}

\newpage
\onecolumngrid
\appendix

\section{Full Evolution Equations}
\label{app:full_evolution_equations}

Using equation~\eqref{eqn:velocity_uniform_soln}, we can eliminate velocity from~\eqref{eqn:theta_S_time_vector_form} and~\eqref{eqn:theta_D_time_vector_form} to obtain evolution equations for the director angles $\theta$ and $\varphi$, giving
\begin{align}
    \begin{split}
        \partial_t \theta & = \frac{K}{\gamma} \Bigl( \partial_z^2 \theta - 2 \tan \varphi \,\partial_z \varphi \,\partial_z \theta \Bigr) - \frac{1-\nu \cos 2\theta}{2\mu} \Bigl( \sigma^{\textrm{el}}_{xz} + \sigma^{\textrm{a}}_{xz} - \bigl\langle \sigma^{\textrm{el}}_{xz} + \sigma^{\textrm{a}}_{xz} \bigr\rangle \Bigr) \\
        & \quad - \frac{1+\nu}{2\mu} \tan \varphi \sin \theta \Bigl( \sigma^{\textrm{el}}_{yz} + \sigma^{\textrm{a}}_{yz} - \bigl\langle \sigma^{\textrm{el}}_{yz} + \sigma^{\textrm{a}}_{yz} \bigr\rangle \Bigr) , 
        \label{eqn:theta_S_full}
    \end{split} \\
    \begin{split}
        \partial_t \varphi & = \frac{K}{\gamma} \Bigl( \partial_z^2\varphi + \sin \varphi \cos \varphi \bigl( \partial_z \theta \bigr)^2 \Bigr) - \frac{1-\nu \cos 2\varphi}{2\mu} \cos \theta \Bigl( \sigma^{\textrm{el}}_{yz} + \sigma^{\textrm{a}}_{yz} - \bigl\langle \sigma^{\textrm{el}}_{yz} + \sigma^{\textrm{a}}_{yz} \bigr\rangle \Bigr) \\
        & \quad - \frac{\nu}{4\mu} \sin 2\varphi \sin 2\theta \Bigl( \sigma^{\textrm{el}}_{xz} + \sigma^{\textrm{a}}_{xz} - \bigl\langle \sigma^{\textrm{el}}_{xz} + \sigma^{\textrm{a}}_{xz} \bigr\rangle \Bigr) ,  
        \label{eqn:theta_D_full}
    \end{split} 
\end{align}
where 
\begin{align}
    \begin{split}
       \sigma^{\textrm{el}}_{xz} + \sigma^{\textrm{a}}_{xz} & = - \frac{K(1-\nu \cos 2\theta)}{2} \Bigl( \cos^2 \varphi \,\partial_z^2 \theta - \sin 2\varphi \,\partial_z \varphi \,\partial_z \theta \Bigr) \\
       & \quad - \frac{K\nu}{4} \sin 2\varphi \sin 2\theta \Bigl( \partial_z^2 \varphi + \sin \varphi \cos \varphi \bigl( \partial_z \theta \bigr)^2 \Bigr) - \zeta \cos^2 \varphi \sin \theta \cos \theta , 
    \end{split} \\
    \begin{split}
        \sigma^{\textrm{el}}_{yz} + \sigma^{\textrm{a}}_{yz} & = - \frac{K(1-\nu \cos 2\varphi) \cos \theta}{2} \Bigl( \partial_z^2 \varphi + \sin \varphi \cos \varphi \bigl( \partial_z \theta \bigr)^2 \Bigr) \\
       & \quad - \frac{K(1+\nu)}{2} \sin \varphi \sin \theta \Bigl( \cos \varphi \,\partial_z^2 \theta - 2 \sin \varphi \,\partial_z \varphi \,\partial_z \theta \Bigr) - \zeta \sin \varphi \cos \varphi \cos \theta .
    \end{split}
\end{align}

\section{Growth Rate Calculations}
\label{app:B}
%%%
\subsection{Isotropic D-mode Growth Rate}
\label{app:isotropic}
We calculate the exponential growth rate of a D-mode perturbation to the ground state, $\vb{n}=\vb{e}_z$, which is isotropic within the cell plane. In this case, both $\theta$ and $\varphi$ are infinitesimal, meaning we consider the stability operator given by \eqref{eqn:linear_theta_operator} (recall that the equations for $\theta$ and $\varphi$ both linearise to the same equation in this case). As already established, the D-mode eigenfunction at threshold is
\begin{equation}
    \theta_0\propto\; 1 - \cos \frac{2\pi z}{d},\\
\end{equation}
Next, we expand \eqref{eqn:linear_theta_operator} in powers of $\zeta-\zeta_\textrm{th}$ to give
\begin{equation}
    \mathcal{L}_1\theta_0 =\;\frac{(1-\nu)(\zeta-\zeta_{\textrm{th}})}{2\mu}\Bigl(\theta_0-\left\langle\theta_0\right\rangle\Bigr).
\end{equation}
Next, we substitute these results into equation~\eqref{eqn:eigenvalue_equation_full} to give
\begin{equation}
\begin{split}
     \lambda_\textrm{D}=& \frac{(\zeta-\zeta_{\text{th}})(1-\nu)\left(\left\langle \theta_0^2 \right\rangle-\left\langle \theta_0 \right\rangle^2\right)}{2\mu\left(\left\langle \theta_0^2 \right\rangle + \frac{\gamma(1-\nu)^2}{4\mu} \left\langle \theta_0 \right\rangle^2\right)} \\
    =& \frac{1-\nu}{6\mu+\gamma(1-\nu)^2}\bigl(\zeta-\zeta_{\text{th}}\bigr).
\end{split}
\end{equation}
This is the result given in equation~\eqref{eqn:D_mode_eigenvalue} in the main text.

\subsection{Anisotropic D-mode Growth Rates}
\label{app:anisotropic}
In this section, we calculate the exponential growth rates of D-mode perturbations on top of an anisotropic steady state due to an S-mode established in the spontaneously chosen direction, $\vb{e}_x$. The S-mode is denoted by $\theta^*(z)$ and is the solution to equation~\eqref{eqn:full_non_linear_single_theta}. We consider the two cases: a D-mode perturbation to $\varphi$, $\theta=\theta^*(z)$ and $\varphi=\delta\varphi_{\textrm{D}}(z,t)$, and a D-mode perturbation to $\theta$, $\theta=\theta^*(z)+\delta\theta_{\textrm{D}}(z,t)$. In the small angle regime close to threshold, these perturbations have leading order contributions along $\vb{e}_y$ and $\vb{e}_x$ respectively, thus we interpret them as perpendicular and parallel perturbations to the direction of anisotropy.
%%%
\subsubsection{Perpendicular Growth Rate}
\label{app:anisotropic_perp}
We substitute $\theta=\theta^*(z)$ and $\varphi=\delta\varphi_{\textrm{D}}(z,t)$ into equation~\eqref{eqn:theta_D_full} and linearise about $\delta\varphi_{\textrm{D}}$, giving
    \begin{equation}
    \begin{split}
        \partial_t\delta\varphi_{\textrm{D}}=&\frac{K}{\gamma}\partial_z^2\delta\varphi_{\textrm{D}}+\frac{K}{\gamma}\left(\partial_z\theta^*\right)^2\,\delta\varphi_{\textrm{D}}\\
        &+\frac{K(1-\nu)^2}{4\mu}\biggl(\cos^2\theta^*\left(\partial_z^2\delta\varphi_{\textrm{D}}+\delta\varphi_{\textrm{D}} \,\left(\partial_z\theta^*\right)^2\right)-\cos\theta^*\left\langle\cos\theta^*\left(\partial_z^2\delta\varphi_{\textrm{D}}+\delta\varphi_{\textrm{D}} \,\left(\partial_z\theta^*\right)^2\right)\right\rangle\biggr)\\
        &+\frac{K\bigl(1-\nu^2\bigr)}{4\mu}\biggl(\cos\theta^*\sin\theta^* \,\partial_z^2\theta^*\,\delta\varphi_{\textrm{D}}-\cos\theta^*\left\langle\sin\theta^* \,\partial_z^2\theta^*\,\delta\varphi_{\textrm{D}}\right\rangle\biggr)\\
        &+\frac{K\nu}{4\mu}\biggl(\sin2\theta^*\partial_z^2\theta^*\bigl(1-\nu\cos2\theta^*\bigr)-\sin2\theta^*\left\langle\partial_z^2\theta^*\bigl(1-\nu\cos2\theta^*\bigr)\right\rangle\biggr)\delta\varphi_{\textrm{D}}\\
        &+\frac{(1-\nu)\zeta}{2\mu} \biggl(\cos^2\theta^* \,\delta\varphi_{\textrm{D}}-\cos\theta^*\left\langle\cos\theta^* \,\delta\varphi_{\textrm{D}}\right\rangle\biggr)+\frac{\nu\zeta}{4\mu} \biggl(\sin^22\theta^*-\sin2\theta^*\left\langle\sin2\theta^*\right\rangle\biggr)\delta\varphi_{\textrm{D}}.
        \label{eqn:perp_stability_operator}
    \end{split}
\end{equation}
The right-hand side is the stability operator, $\mathcal{L}$, of $\delta\varphi_{\textrm{D}}$. In the limit $\zeta\to\zeta_\textrm{th}$, $\theta^*\to 0$ and the right-hand side of equation~\eqref{eqn:perp_stability_operator} reduces to the form given in equation~\eqref{eqn:threshold_operator}, meaning that we can apply the perturbation theory outlined in \S\ref{sec:methods}. Close to threshold, $\theta^*=A_\textrm{S}^*\sin\frac{2\pi z}{d}+\text{higher order terms}$, where $A_\textrm{S}^*\sim (\zeta-\zeta_\textrm{th})^{\frac{1}{2}}$. Therefore, upon expanding the stability operator in $\theta^*$, we obtain the expansion $\mathcal{L}=\mathcal{L}_0+\mathcal{L}_1+\dots$, with terms $\order{\theta^*}^2$ contributing to $\mathcal{L}_1$. 
\begin{equation}
    \begin{split}
        \mathcal{L}_1\delta\varphi_{\textrm{D}}=&\frac{K}{\gamma}\left(\partial_z\theta^*\right)^2\,\delta\varphi_{\textrm{D}}+\frac{K(1-\nu)^2}{4\mu}\Bigl(\left(\partial_z\theta^*\right)^2\delta\varphi_{\textrm{D}}-\left\langle\left(\partial_z\theta^*\right)^2\delta\varphi_{\textrm{D}}\right\rangle\Bigr)\\
        &+\frac{K\bigl(1-\nu^2\bigr)}{4\mu}\Bigl(\theta^*\partial_z^2\theta^*\delta\varphi_{\textrm{D}}-\left\langle\theta^*\partial_z^2\theta^*\delta\varphi_{\textrm{D}}\right\rangle\Bigr)+\frac{K\nu(1-\nu)}{2\mu}\theta^*\Bigl(\partial_z^2\theta^*-\left\langle\partial_z^2\theta^*\right\rangle\Bigr)\delta\varphi_{\textrm{D}}\\
        &-\frac{K(1-\nu)^2}{8\mu}\left(\theta^*\right)^2\Bigl(\partial_z^2\delta\varphi_{\textrm{D}}-\left\langle\partial_z^2\delta\varphi_{\textrm{D}}\right\rangle\Bigr)-\frac{K(1-\nu)^2}{8\mu}\left(\left(\theta^*\right)^2\partial_z^2\delta\varphi_{\textrm{D}}-\left\langle\left(\theta^*\right)^2\partial_z^2\delta\varphi_{\textrm{D}}\right\rangle\right)\\
        &+\frac{(\zeta-\zeta_\textrm{th})(1-\nu)}{2\mu}\Bigl(\delta\varphi_{\textrm{D}}-\left\langle\delta\varphi_{\textrm{D}}\right\rangle\Bigr)-\frac{\zeta_\textrm{th}(1-\nu)}{4\mu}\left(\theta^*\right)^2\Bigl(\delta\varphi_{\textrm{D}}-\left\langle\delta\varphi_{\textrm{D}}\right\rangle\Bigr)\\
        &-\frac{\zeta_\textrm{th}(1-\nu)}{4\mu}\left(\left(\theta^*\right)^2\delta\varphi_{\textrm{D}}-\left\langle\left(\theta^*\right)^2\delta\varphi_{\textrm{D}}\right\rangle\right)+\frac{\zeta_\textrm{th}\nu}{\mu}\Bigl(\theta^*-\left\langle\theta^*\right\rangle\Bigr)\theta^*\delta\varphi_{\textrm{D}}.
    \end{split}
\end{equation}
Only the leading order term of $\theta^*$ contributes to this order. Thus, to calculate the growth rate, we evaluate equation~\eqref{eqn:eigenvalue_equation_full} with $\delta\varphi_{\textrm{D},0}\propto1-\cos\frac{2\pi z}{d}$ and $\theta^*=A_\textrm{S}^*\sin\frac{2\pi z}{d}$. The result is given by equation~\eqref{eqn:perp_growth_rate} in the main text.

%%%

\subsubsection{Parallel Growth Rate}
\label{app:anisotropic_parr}
The calculation is the essentially the same as that given in Appendix~\ref{app:anisotropic_perp}, in which more detail is given. We substitute $\theta=\theta^*(z)+\delta\theta_{\textrm{D}}(z,t)$ and $\varphi=0$ into equation \eqref{eqn:theta_S_full} and linearise about $\delta\theta_{\textrm{D}}$, giving
    \begin{equation}
    \begin{split}
         \partial_t\delta\theta_{\textrm{D}}=&\frac{K}{\gamma}\partial_z^2\delta\theta_{\textrm{D}}+\frac{K\nu}{2\mu}\bigl(1-\nu\cos2\theta^*\bigr)\Bigl(\sin2\theta^*\partial_z^2\theta^*\,\delta\theta_{\textrm{D}}-\left\langle \sin2\theta^*\partial_z^2\theta^*\,\delta\theta_{\textrm{D}}\right\rangle\Bigr)\\
         &+\frac{K}{4\mu}\bigl(1-\nu\cos2\theta^*\bigr)\Bigl(\bigl(1-\nu\cos2\theta^*\bigr)\,\partial_z^2\delta\theta_{\textrm{D}}-\left\langle \bigl(1-\nu\cos2\theta^*\bigr)\,\partial_z^2\delta\theta_{\textrm{D}}\right\rangle\Bigr)\\
         &+\frac{K\nu}{2\mu}\Bigl(\sin2\theta^*\partial_z^2\theta^*\bigl(1-\nu\cos2\theta^*\bigr)-\sin2\theta^*\left\langle \partial_z^2\theta^*\bigl(1-\nu\cos2\theta^*\bigr)\right\rangle\Bigr)\,\delta\theta_{\textrm{D}}\\
         &+\frac{\zeta}{2\mu}\bigl(1-\nu\cos2\theta^*\bigr)\Bigl(\cos2\theta^*\,\delta\theta_{\textrm{D}}-\left\langle \cos2\theta^*\,\delta\theta_{\textrm{D}}\right\rangle\Bigr)+\frac{\nu\zeta}{2\mu}\Bigl(\sin2\theta^* \sin2\theta^*-\sin2\theta^*\left\langle\sin2\theta^*\right\rangle\Bigr)\,\delta\theta_{\textrm{D}}.
         \label{eqn:parr_stability_operator}
    \end{split}
\end{equation}
From this, we expand in $\theta^*$ to obtain
\begin{equation}
    \begin{split}
         \mathcal{L}_1\delta\theta_{\textrm{D}}=&\frac{K\nu(1-\nu)}{\mu}\Bigl(\theta^*\partial_z^2\theta^*\,\delta\theta_{\textrm{D}}-\bigl\langle \theta^*\partial_z^2\theta^*\,\delta\theta_{\textrm{D}}\bigr\rangle\Bigr)+\frac{K\nu(1-\nu)}{2\mu}\left(\left(\theta^*\right)^2\,\partial_z^2\delta\theta_{\textrm{D}}-\left\langle\left(\theta^*\right)^2\,\partial_z^2\delta\theta_{\textrm{D}}\right\rangle\right)\\
         &+\frac{K\nu(1-\nu)}{2\mu}\left(\theta^*\right)^2\Bigl(\partial_z^2\delta\theta_{\textrm{D}}-\bigl\langle\partial_z^2\delta\theta_{\textrm{D}}\bigr\rangle\Bigr)+\frac{K\nu(1-\nu)}{\mu}\theta^*\Bigl(\partial_z^2\theta^*-\bigl\langle\partial_z^2\theta^*\bigr\rangle\Bigr)\delta\theta_{\textrm{D}}\\
         &+\frac{(1-\nu)(\zeta-\zeta_\textrm{th})}{2\mu}\Bigl(\delta\theta_{\textrm{D}}-\bigl\langle\delta\theta_{\textrm{D}}\bigr\rangle\Bigr)-\frac{\zeta_\textrm{th}(1-\nu)}{\mu}\left(\left(\theta^*\right)^2\delta\theta_{\textrm{D}}-\left\langle\left(\theta^*\right)^2\delta\theta_{\textrm{D}}\right\rangle\right)\\
         &+\frac{\zeta_\textrm{th}\nu}{\mu}\left(\theta^*\right)^2\Bigl(\delta\theta_{\textrm{D}}-\bigl\langle\delta\theta_{\textrm{D}}\bigr\rangle\Bigr)+\frac{2\nu\zeta_\textrm{th}}{\mu}\theta^*\Bigl(\theta^*-\bigl\langle\theta^*\bigr\rangle\Bigr)\delta\theta_{\textrm{D}}.
         \label{eqn:parr_stability_operator}
    \end{split}
\end{equation}
We again evaluate equation~\eqref{eqn:eigenvalue_equation_full} with $\delta\theta_{\textrm{D},0}\propto1-\cos\frac{2\pi z}{d}$ and $\theta^*=A_\textrm{S}^*\sin\frac{2\pi z}{d}$. The result is given by equation~\eqref{eqn:parr_growth_rate} in the main text.

%%%

\section{Amplitude Evolution Equations}
\label{app:steady_state}
\subsection{Calculation of the S-mode Amplitude Evolution Equation}
\label{app:S-mode_evolution_calc}
We first expand equation~\eqref{eqn:theta_S_full} up to combined cubic order in $\theta$ and $\varphi$, giving
    \begin{equation}
    \begin{split}
         \partial_t\theta=&\frac{K}{\gamma}\partial_z^2\theta+\frac{K(1-\nu)^2}{4\mu}\biggl(\partial_z^2\theta-\left\langle\partial_z^2\theta\right\rangle\Bigr)+\frac{K\nu(1-\nu)}{\mu}\Bigl(\theta^2\partial_z^2\theta-\frac{1}{2}\left\langle\theta^2\partial_z^2\theta\right\rangle-\frac{1}{2}\theta^2\left\langle\partial_z^2\theta\right\rangle\biggr)\\
         &+\frac{K\nu(1-\nu)}{2\mu}\Bigl(\varphi\theta\partial_z^2\varphi-\left\langle\varphi\theta\partial_z^2\varphi\right\rangle\Bigr)+\frac{K(1-\nu)(1+\nu)}{4\mu}\Bigl(\varphi\theta\partial_z^2\varphi-\varphi\theta\left\langle\partial_z^2\varphi\right\rangle\Bigr)-2\frac{K}{\gamma}\varphi\,\partial_z\varphi\,\partial_z\theta\\
         &-\frac{K(1-\nu)^2}{4\mu}\Bigl(\varphi^2\partial_z^2\theta-\left\langle\varphi^2\partial_z^2\theta\right\rangle\Bigr)-\frac{K(1-\nu)^2}{2\mu}\Bigl(\varphi\partial_z\varphi\partial_z\theta-\left\langle\varphi\partial_z\varphi\partial_z\theta\right\rangle\Bigr)-\frac{\zeta(1-\nu)}{2\mu}\Bigl(\varphi^2\theta-\left\langle\varphi^2\theta\right\rangle\Bigr)\\
         &+\frac{\zeta(1-\nu)}{2\mu}\Bigl(\theta-\left\langle\theta\right\rangle\Bigr)-\frac{\zeta(1-\nu)}{3\mu}\Bigl(\theta^3-\left\langle\theta^3\right\rangle\Bigr)+\frac{\zeta \nu}{\mu}\Bigl(\theta^3-\theta^2\left\langle\theta\right\rangle\Bigr)+\frac{\zeta(1+\nu)}{2\mu} \Bigl(\theta\varphi^2-\varphi\theta\left\langle\varphi\right\rangle\Bigr).
    \end{split}
\end{equation}
Only $\theta_0$ and $\varphi_0$ contribute to $\mathcal{L}_1\theta_0$, meaning that
\begin{equation}
    \begin{split}
         \mathcal{L}_1\theta_0=&\frac{K\nu(1-\nu)}{\mu}\Bigl(\theta_0^2\partial_z^2\theta_0-\frac{1}{2}\left\langle\theta_0^2\partial_z^2\theta_0\right\rangle-\frac{1}{2}\theta_0^2\left\langle\partial_z^2\theta_0\right\rangle\Bigr)+\frac{K\nu(1-\nu)}{2\mu}\Bigl(\varphi_0\theta_0\partial_z^2\varphi_0-\left\langle\varphi_0\theta_0\partial_z^2\varphi_0\right\rangle\Bigr)\\
         &+\frac{K(1-\nu^2)}{4\mu}\Bigl(\varphi_0\theta_0\partial_z^2\varphi_0-\varphi_0\theta_0\left\langle\partial_z^2\varphi_0\right\rangle\Bigr)-2\frac{K}{\gamma}\varphi_0\,\partial_z\varphi_0\,\partial_z\theta_0\\
         &-\frac{K(1-\nu)^2}{4\mu}\Bigl(\varphi_0^2\partial_z^2\theta_0-\left\langle\varphi_0^2\partial_z^2\theta_0\right\rangle\Bigr)-\frac{K(1-\nu)^2}{2\mu}\Bigl(\varphi_0\partial_z\varphi_0\partial_z\theta_0-\left\langle\varphi_0\partial_z\varphi_0\partial_z\theta_0\right\rangle\Bigr)\\
         &+\frac{(1-\nu)(\zeta-\zeta_\textrm{th})}{2\mu}\Bigl(\theta_0-\left\langle\theta_0\right\rangle\Bigr)-\frac{\zeta_\textrm{th}(1-\nu)}{3\mu}\bigl(\theta_0^3-\left\langle\theta_0^3\right\rangle\bigr)+\frac{\zeta_\textrm{th} \nu}{\mu}\Bigl(\theta_0^3-\theta_0^2\left\langle\theta_0\right\rangle\Bigr)\\
         &-\frac{\zeta_\textrm{th}(1-\nu)}{2\mu}\Bigl(\varphi_0^2\theta_0-\left\langle\varphi_0^2\theta_0\right\rangle\Bigr)+\frac{\zeta_\textrm{th}(1+\nu)}{2\mu} \Bigl(\theta_0\varphi_0^2-\varphi_0\theta_0\left\langle\varphi_0\right\rangle\Bigr).
    \end{split}
\end{equation}
In this case, equation~\eqref{eqn:amp_evol_general}, reduces to
\begin{equation}
    \dv{A_\textrm{S}}{t}=\frac{\langle\psi_\textrm{S}\mathcal{L}_1 \theta_0\rangle}{\langle\psi_\textrm{S}^2\rangle},
    \label{eqn:A_S_reduced}
\end{equation}
After evaluating equation~\eqref{eqn:A_S_reduced}, we obtain the coefficients
\begin{gather}
    \Lambda_1=\frac{(1-4\nu)+\frac{\gamma(1-\nu)^2}{4\mu}(1+2\nu)}{4\biggl(1+\frac{\gamma(1-\nu)^2}{4\mu}\biggr)}\frac{\zeta_\textrm{th}}{\mu},\\
    \Lambda_2=\frac{(1-2\nu)\biggl(\frac{\gamma(1-\nu)^2}{4\mu}-1\biggr)}{16\biggl(1+\frac{\gamma(1-\nu)^2}{4\mu}\biggr)}\frac{\zeta_\textrm{th}}{\mu}.
\end{gather}

%%%
\subsection{Calculation of the D-mode Amplitude Evolution Equation}

Following the same method as in Appendix~\ref{app:S-mode_evolution_calc}, we expand \eqref{eqn:theta_D_full} up to cubic order, giving
    \begin{equation}
    \begin{split}
        \partial_t\varphi=&\frac{K}{\gamma}\partial_z^2\varphi+\frac{K(1-\nu)^2}{4\mu}\Bigl(\partial_z^2\varphi-\left\langle\partial_z^2\varphi\right\rangle\Bigr)+\frac{K\nu(1-\nu)}{2\mu}\Bigl(\varphi\theta\partial_z^2\theta-\varphi\theta\left\langle\partial_z^2\theta\right\rangle\Bigr)+ \frac{K}{\gamma}\varphi(\partial_z\theta)^2\\
        &+\frac{K\nu(1-\nu)}{2\mu}\Bigl(2\varphi^2\partial_z^2\varphi-\left\langle\varphi^2\partial_z^2\varphi\right\rangle-\varphi^2\left\langle\partial_z^2\varphi\right\rangle\Bigr)+\frac{K\bigl(1-\nu^2\bigr)}{4\mu}\Bigl(\varphi\theta \partial_z^2\theta-\left\langle\varphi\theta \partial_z^2\theta\right\rangle\Bigr)\\
        &+\frac{K(1-\nu)^2}{4\mu}\Bigl(\varphi(\partial_z\theta)^2-\left\langle\varphi(\partial_z\theta)^2\right\rangle\Bigr)-\frac{K(1-\nu)^2}{8\mu}\Bigl(2\theta^2\partial_z^2\varphi-\left\langle\theta^2\partial_z^2\varphi\right\rangle-\theta^2\left\langle\partial_z^2\varphi\right\rangle\Bigr)\\
        &+\frac{\zeta(1-\nu)}{2\mu}\Bigl(\varphi-\left\langle\varphi\right\rangle\Bigr)-\frac{\zeta(1-\nu)}{3\mu}\Bigl(\varphi^3-\left\langle\varphi^3\right\rangle\Bigr)+\frac{\zeta\nu}{\mu}\Bigl(\varphi^3-\varphi^2\left\langle\varphi\right\rangle\Bigr)\\
        &+\frac{\zeta\nu}{\mu}\Bigl(\theta^2\varphi-\varphi\theta\left\langle\theta\right\rangle\Bigr)-\frac{\zeta(1-\nu)}{4\mu}\Bigl(2\theta^2\varphi-\left\langle\theta^2\varphi\right\rangle-\theta^2\left\langle\varphi\right\rangle\Bigr),
    \end{split}
\end{equation}
from which we extract
    \begin{equation}
    \begin{split}
        \mathcal{L}_1\varphi_0=&\frac{K}{\gamma}\varphi_0(\partial_z\theta_0)^2+\frac{K\nu(1-\nu)}{2\mu}\Bigl(2\varphi_0^2\partial_z^2\varphi_0-\left\langle\varphi_0^2\partial_z^2\varphi_0\right\rangle-\varphi_0^2\left\langle\partial_z^2\varphi_0\right\rangle\Bigr)\\
        &+\frac{K(1-\nu)^2}{4\mu}\Bigl(\varphi_0(\partial_z\theta_0)^2-\left\langle\varphi_0(\partial_z\theta_0)^2\right\rangle\Bigr)-\frac{K(1-\nu)^2}{8\mu}\Bigl(2\theta_0^2\partial_z^2\varphi_0-\left\langle\theta_0^2\partial_z^2\varphi_0\right\rangle-\theta_0^2\left\langle\partial_z^2\varphi_0\right\rangle\Bigr)\\
        &+\frac{K\nu(1-\nu)}{2\mu}\Bigl(\varphi_0\theta_0\partial_z^2\theta_0-\varphi_0\theta_0\left\langle\partial_z^2\theta_0\right\rangle\Bigr)+\frac{K(1-\nu^2)}{4\mu}\Bigl(\varphi_0\theta_0 \partial_z^2\theta_0-\left\langle\varphi_0\theta_0 \partial_z^2\theta_0\right\rangle\Bigr)\\
        &+\frac{(1-\nu)(\zeta-\zeta_\textrm{th})}{2\mu}\Bigl(\varphi_0-\left\langle\varphi_0\right\rangle\Bigr)-\frac{\zeta_\textrm{th}(1-\nu)}{3\mu}\Bigl(\varphi_0^3-\left\langle\varphi_0^3\right\rangle\Bigr)+\frac{\zeta_\textrm{th}\nu}{\mu}\Bigl(\varphi_0^3-\varphi_0^2\left\langle\varphi_0\right\rangle\Bigr)\\
        &+\frac{\zeta_\textrm{th}\nu}{\mu}\Bigl(\theta_0^2\varphi_0-\varphi_0\theta_0\left\langle\theta_0\right\rangle\Bigr)-\frac{\zeta_\textrm{th}(1-\nu)}{4\mu}\Bigl(2\theta_0^2\varphi_0-\left\langle\theta_0^2\varphi_0\right\rangle-\theta_0^2\left\langle\varphi_0\right\rangle\Bigr).
    \end{split}
\end{equation}
In this case, equation~\eqref{eqn:amp_evol_general} is evaluated as
\begin{equation}
    \dv{A_\textrm{D}}{t}=\frac{\langle\psi_\textrm{D}\mathcal{L}_1 \varphi_0\rangle+\frac{\gamma(1-\nu)^2}{4\mu}\langle\psi_\textrm{D}\rangle\langle\mathcal{L}_1 \varphi_0\rangle}{\langle\psi_\textrm{D}^2\rangle+\frac{\gamma(1-\nu)^2}{4\mu}\langle\psi_\textrm{D}\rangle^2},
\end{equation}
from which we obtain the coefficients
\begin{gather}
    \Lambda_3=\frac{(3+2\nu)+\frac{3\gamma(1-\nu)^2}{4\mu}}{4\Bigl(3+\frac{\gamma(1-\nu)^2}{2\mu}\Bigr)\Bigl(1+\frac{\gamma(1-\nu)^2}{4\mu}\Bigr)}\frac{\zeta_\textrm{th}}{\mu},\\
    \Lambda_4=\frac{5(1-4\nu)+\frac{\gamma(1-\nu)^2}{4\mu}(5-6\nu)}{16\Bigl(3+\frac{\gamma(1-\nu)^2}{2\mu}\Bigr)\Bigl(1+\frac{\gamma(1-\nu)^2}{4\mu}\Bigr)}\frac{\zeta_\textrm{th}}{\mu}.
\end{gather}


\begin{thebibliography}{99}
\bibitem{RamaswarmyMech2010} S. Ramaswamy, The Mechanics and Statistics of Active Matter, Annu. Rev. Condens. Matter Phys. \textbf{1}, 323-345 (2010). \doi{10.1146/annurev-conmatphys-070909-104101}
\bibitem{marchetti2013hydrodynamics} M.C. Marchetti, J.F. Joanny, S. Ramaswamy, T.B. Liverpool, J. Prost, M. Rao, and R.A. Simha, Hydrodynamics of soft active matter, Rev. Mod. Phys. \textbf{85}, 1143 (2013). \doi{10.1103/RevModPhys.85.1143}
\bibitem{finlayson1969} B.A. Finlayson and L.E. Scriven, Convective instability by active stress, Proc. R. Soc. Lond. A {\bf 310}, 183 (1969). \doi{10.1098/rspa.1969.0071}
\bibitem{doostmohammadi2018active} A. Doostmohammadi, J. Ignés-Mullol, J.M Yeomans, and F. Sagués, Active nematics, Nat. Commun. \textbf{9}, 1-13 (2018). \doi{10.1038/s41467-018-05666-8}
\bibitem{duclos2017} G. Duclos, C. Erlenk\"amper, J-F. Joanny, and P. Silberzan, Topological defects in confined populations of spindle-shaped cells, Nat. Phys. {\bf 13}, 58 (2017). \doi{10.1038/nphys3876}
\bibitem{meacock2021} O.J. Meacock, A. Doostmohammadi, K.R. Foster, J.M. Yeomans, and W.M. Durham, Bacteria solve the problem of crowding by moving slowly, Nat. Phys. {\bf 17}, 205 (2021). \doi{10.1038/s41567-020-01070-6}
\bibitem{copenhagen2021} K. Copenhagen, R. Alert, N.S. Wingreen, and J.W. Shaevitz, Topological defects promote layer formation in {\it Myxococcus xanthus} colonies, Nat. Phys. {\bf 17}, 211 (2021). \doi{10.1038/s41567-020-01056-4}
\bibitem{saw2017topological} T.B. Saw, A. Doostmohammadi, V. Nier, L. Kocgozlu, S.P. Thampi, Y. Toyama, P. Marcq, C.T. Lim, J.M. Yeomans, and B. Ladoux, Topological defects in epithelia govern cell death and extrusion, Nature \textbf{544}, 212-216 (2017). \doi{10.1038/nature21718}
\bibitem{doostmohammadi2022} A. Doostmohammadi and B. Ladoux, Physics of liquid crystals in cell biology, Trends Cell Bio. {\bf 32}, 140 (2022). \doi{10.1016/j.tcb.2021.09.012}
\bibitem{wensink2012meso} H.H. Wensink, J. Dunkel, S. Heidenreich, K. Drescher, R.E. Goldstein, H. Löwen, and J.M. Yeomans, Meso-scale turbulence in living fluids, Proc. Natl. Acad. Sci. U.S.A. \textbf{109}, 14308-14313 (2012). \doi{10.1073/pnas.1202032109}
\bibitem{wioland2016} H. Wioland, E. Lushi, and R.E. Goldstein, Directed collective motion of bacteria under channel confinement, New J. Phys. {\bf 18}, 075002 (2016). \doi{10.1088/1367-2630/18/7/075002}
\bibitem{dunkel2013fluid}J. Dunkel, S. Heidenreich, K. Drescher, H.H. Wensink, M. Bär, and R.E. Goldstein, Fluid dynamics of bacterial turbulence, Phys. Rev. Lett. \textbf{110}, 228102 (2013). \doi{10.1103/PhysRevLett.110.228102}
\bibitem{sanchez2012spontaneous} T. Sanchez, D.T.N Chen, S.J. DeCamp, M. Heymann, and Z. Dogic, Spontaneous motion in hierarchically assembled active matter, Nature \textbf{491}, 431-434 (2012). \doi{10.1038/nature11591}
\bibitem{narayan2007} V. Narayan, S. Ramaswamy, and N. Menon, Long-Lived Giant Number Fluctuations in a Swarming Granular Nematic, Science {\bf 317}, 105 (2007). \doi{10.1126/science.1140414}
\bibitem{kumar2014flocking} N. Kumar, H. Soni, S. Ramaswamy, and A.K. Sood, Flocking at a distance in active granular matter, Nat. Commun. \textbf{5}, 1-9 (2014). \doi{10.1038/ncomms5688}
\bibitem{mclennan2012multiscale} R. McLennan, L. Dyson, K.W. Prather, J.A. Morrison, R.E. Baker, P.K. Maini, and P.M. Kulesa, Multiscale mechanisms of cell migration during development: theory and experiment, Development \textbf{139}, 2935-2944 (2012). \doi{10.1242/dev.081471}
\bibitem{poujade2007collective} M. Poujade, E. Grasland-Mongrain, A. Hertzog, J. Jouanneau, P. Chavrier, B. Ladoux, A. Buguin, and P. Silberzan, Collective migration of an epithelial monolayer in response to a model wound, Proc. Natl. Acad. Sci. U.S.A. \textbf{104}, 15988-15993 (2007). \doi{10.1073/pnas.0705062104}
\bibitem{houston2023} A.J.H. Houston and G.P. Alexander, Active nematic multipoles: Flow responses and the dynamics of defects and colloids, Front. Phys. {\bf 11}, 1110244 (2023). \doi{10.3389/fphy.2023.1110244}
\bibitem{houston2023cog} A.J.H. Houston and G.P. Alexander, Colloids in Two-Dimensional Active Nematics: Conformal Cogs and Controllable Spontaneous Rotation, arxiv:2307.05247 [cond-mat.soft] (2023). \doi{10.48550/arXiv.2307.05247}
\bibitem{ray2023} S. Ray, J. Zhang, and Z. Dogic, Rectified Rotational Dynamics of Mobile Inclusions in Two-Dimensional Active Nematics, Phys. Rev. Lett. {\bf 130}, 238301 (2023). \doi{10.1103/PhysRevLett.130.238301}
\bibitem{beris1994thermodynamics} A.N. Beris, and B.J. Edwards, {\sl Thermodynamics of flowing systems: with internal microstructure}, (Oxford University Press, Oxford, 1994).
\bibitem{deGennesProst} P.G. de Gennes and J. Prost, {\sl The Physics of Liquid Crystals}, (Oxford University Press, Oxford, 1993).
\doi{10.1103/PhysRevLett.89.058101}
\bibitem{simha2002hydrodynamic} R.A. Simha, and S. Ramaswamy, Hydrodynamic fluctuations and instabilities in ordered suspensions of self-propelled particles, Phys. Rev Lett. \textbf{89}, 058101 (2002).
\bibitem{alert2022active} R. Alert, J. Casademunt, and J.F. Joanny, Active turbulence, Annu. Rev. Condens. Matter Phys. \textbf{13}, 143-170 (2022). \doi{10.1146/annurev-conmatphys-082321-035957}
\bibitem{voituriez2005} R. Voituriez, J.F. Joanny, and J. Prost, Spontaneous flow transition in active polar gels, EPL {\bf 70}, 404 (2005). \doi{10.1209/epl/i2004-10501-2}
\bibitem{marenduzzo2007} D. Marenduzzo, E. Orlandini, M.E. Cates, and J.M. Yeomans, Steady-state hydrodynamic instabilities of active liquid crystals: Hybrid lattice Boltzmann simulations, Phys. Rev. E {\bf 76}, 031921 (2007). \doi{10.1103/PhysRevE.76.031921}
\bibitem{duclos2018} G. Duclos, C. Blanch-Mercader, V. Yashunsky, G. Salbreux, J.F. Joanny, J. Prost, and P. Silberzan, Spontaneous shear flow in confined cellular nematics, Nat. Phys. {\bf 14}, 728 (2018). \doi{10.1038/s41567-018-0099-7}
\bibitem{thampi2022channel} S.P. Thampi. Channel Confined Active Nematics,  Curr. Op. Coll. Interface Sci. \textbf{61}, 101613 (2022).
\doi{10.1016/j.cocis.2022.101613}
\bibitem{opathalage2019self} A. Opathalage, M.M. Norton, M.P.N. Juniper, B. Langeslay, S. A. Aghvami, S. Fraden, and Z. Dogic, Self-organized dynamics and the transition to turbulence of confined active nematics, Proc. Natl. Acad. Sci. U.S.A. \textbf{116}, 4788-4797 (2019). \doi{10.1073/pnas.1816733116}
\bibitem{zumdieck2008} A. Zumdieck, R. Voituriez, J. Prost, and J.F. Joanny, Spontaneous flow of active polar gels in undulated channels, Faraday Discuss. {\bf 139}, 369 (2008). \doi{10.1039/B716934E}
\bibitem{rorai2021active} C. Rorai, F. Toschi, and I. Pagonabarraga, Active nematic flows confined in a two-dimensional channel with hybrid alignment at the walls: A unified picture, Phys. Rev. Fluids \textbf{6}, 113302 (2021). \doi{10.1103/PhysRevFluids.6.113302}
\bibitem{edwards2009} S.A. Edwards and J.M. Yeomans, Spontaneous flow states in active nematics: A unified picture, EPL {\bf 85}, 18008 (2009). \doi{10.1209/0295-5075/85/18008}
\bibitem{samui2021flow} A. Samui, J.M. Yeomans, and S.P. Thampi, Flow transitions and length scales of a channel-confined active nematic, Soft Matter \textbf{17}, 10640-10648 (2021). \doi{10.1039/d1sm01434j}
\bibitem{furthauer2012} S. F\"urthauer, M. Neef, S.W. Grill, K. Kruse, and F. J\"ulicher, The Taylor–Couette motor: spontaneous flows of active polar fluids between two coaxial cylinders, New J. Phys. {\bf 14}, 023001 (2012). \doi{10.1088/1367-2630/14/2/023001}
\bibitem{ravnik2013} M. Ravnik and J.M. Yeomans, Confined Active Nematic Flow in Cylindrical Capillaries, Phys. Rev. Lett. {\bf 110}, 026001 (2013). \doi{10.1103/PhysRevLett.110.026001}
\bibitem{shendruk2017dancing} T. Shendruk, A. Doostmohammadi, K. Thijssen, and J.M. Yeomans, Dancing disclinations in confined active nematics, Soft Matter \textbf{13}, 3853-3862 (2017). \doi{10.1039/c6sm02310j}
\bibitem{doostmohammadi2017} A. Doostmohammadi, T.N. Shendruk, K. Thijssen, and J.M. Yeomans, Onset of meso-scale turbulence in active nematics, Nat. Commun. {\bf 8}, 15326 (2017). \doi{10.1038/ncomms15326}
\bibitem{chen2018dynamics} S. Chen, P. Gao, and T. Gao, Dynamics and structure of an apolar active suspension in an annulus, J. Fluid Mech. \textbf{835}, 393-405 (2018). \doi{10.1017/jfm.2017.759}
\bibitem{hardouin2019reconfigurable} J. Hardoüin, R. Hughes, A. Doostmohammadi, J. Laurent, T. Lopez-Leon, J.M. Yeomans, J. Ignés-Mullol, and F. Sagués, Reconfigurable flows and defect landscape of confined active nematics, Commun. Phys. \textbf{2}, 1-9 (2019). \doi{10.1038/s42005-019-0221-x}
\bibitem{chandragiri2019active} S. Chandragiri, A. Doostmohammadi, J.M. Yeomans, and S.P. Thampi, Active transport in a channel: stabilisation by flow or thermodynamics, Soft Matter \textbf{15}, 1597-1604 (2019). \doi{10.1039/C8SM02103A}
\bibitem{wu2017transition} K.T. Wu, J.B. Hishamunda, D.T.N Chen, S.J. DeCamp, Y.W. Chang, A. Fernández-Nieves, S. Fraden, and Z. Dogic, Transition from turbulent to coherent flows in confined three-dimensional active fluids, Science \textbf{355}, eaal1979 (2017). \doi{10.1126/science.aal1979}
\bibitem{shendruk2018twist} T.N. Shendruk, K. Thijssen, J.M. Yeomans, and A. Doostmohammadi, Twist-induced crossover from two-dimensional to three-dimensional turbulence in active nematics, Phys. Rev. E. \textbf{98}, 010601 (2018). \doi{10.1103/PhysRevE.98.010601}
\bibitem{chandragiri2020} S. Chandragiri, A. Doostmohammadi, J.M. Yeomans, and S.P. Thampi, Flow States and Transitions of an Active Nematic in a Three-Dimensional Channel, Phys. Rev. Lett. {\bf 125}, 148002 (2020). \doi{10.1103/PhysRevLett.125.148002}
\bibitem{chandrarkar2020} P. Chandrarkar, M. Varghese, S.A. Aghvami, A. Baskaran, Z. Dogic, and G. Duclos, Confinement Controls the Bend Instability of Three-Dimensional Active Liquid Crystals, Phys. Rev. Lett. {\bf 125}, 257801 (2020). \doi{10.1103/PhysRevLett.125.257801}
\bibitem{varghese2020} M. Varghese, A. Baskaran, M.F. Hagen, and A. Baskaran, Confinement-Induced Self-Pumping in 3D Active Fluids, Phys. Rev. Lett. {\bf 125}, 268003 (2020). \doi{10.1103/PhysRevLett.125.268003}
\bibitem{strubing2020} T. Str\"ubing, A. Khosravanizadeh, A. Vilfan, E. Bodenschatz, R. Golestanian, and I. Guido, Wrinkling instability in 3D active nematics, Nano Lett. {\bf 20}, 6281 (2020). \doi{10.1021/acs.nanolett.0c01546}
\bibitem{fan2021effects} Y. Fan, K.T. Wu, S.A. Aghvami, S. Fraden, and K.S. Breuer, Effects of confinement on the dynamics and correlation scales in kinesin-microtubule active fluids, Phys. Rev. E \textbf{104}, 034601 (2021). \doi{10.1103/PhysRevE.104.034601}
\bibitem{keogh2022helical} R.R. Keogh, S. Chandragiri, B. Loewe, T. Ala-Nissila, S.P. Thampi, and T.N. Shendruk, Helical flow states in active nematics, Phys. Rev. E \textbf{106}, L012602 (2022). \doi{10.1103/PhysRevE.106.L012602}
\bibitem{kruger2017lattice} T. Krüger, H. Kusumaatmaja, A. Kuzmin, O. Shardt,  G. Silva \& E. Viggen,  The lattice Boltzmann method. Springer International Publishing. \textbf{10}, 4-15 (2017)
\bibitem{Chandrakar2020bend}Chandrakar, P., Varghese, M., Aghvami, S., Baskaran, A., Dogic, Z. \& Duclos, G. Confinement Controls the Bend Instability of Three-Dimensional Active Liquid Crystals.  Phys. Rev. Lett. \textbf{125}, 257801 (2020). \doi{10.1103/PhysRevLett.125.257801}
\bibitem{duclos2020topological}G. Duclos, R. Adkins, D. Banerjee, M.S.E Peterson, M. Varghese, I. Kolvin, A. Baskaran, R.A. Pelcovits, T.R. Powers, A. Baskaran, and others. Topological structure and dynamics of three-dimensional active nematics, Science \textbf{367}, 1120-1124 (2020). \doi{10.1126/science.aaz4547}
\bibitem{houston2022defect}A.J.H. Houston, and G.P. Alexander. Defect loops in three-dimensional active nematics as active multipoles, Phys. Rev. E.
\doi{10.1103/PhysRevE.105.L062601}
\bibitem{MachonAlexander2014knots} T. Machon and G.P. Alexander, Knotted Defects in Nematic Liquid Crystals, Phys. Rev. Lett. \textbf{113}, 027801 (2014). \doi{10.1103/PhysRevLett.113.027801}
\bibitem{tkalec2011knots} U. Tkalec et al., Reconfigurable Knots and Links in Chiral Nematic Colloids, Science \textbf{333}, 62-65 (2011). \doi{10.1126/science.1205705}
\bibitem{Martinez2014knots} A. Martinez, M. Ravnik, B. Lucero et al. Mutually tangled colloidal knots and induced defect loops in nematic fields. Nature Materials \textbf{13}, 258–263 (2014). \doi{10.1038/nmat3840}
\bibitem{self-assembly} H.-S. Park, S.-W. Kang, L. Tortora, Y. Nastishin, D. Finotello, S. Kumar, and O. D. Lavrentovich
The Journal of Physical Chemistry B \textbf{112} (51), 16307-16319 (2008)
\doi{10.1021/jp804767z}
\bibitem{multipoles} B. Senyuk, J. Aplinc, M. Ravnik, et al. High-order elastic multipoles as colloidal atoms. Nat. Commun. \textbf{10}, 1825 (2019). \doi{10.1038/s41467-019-09777-8}
\bibitem{hexadecapolar} B. Senyuk, O. Puls, O. Tovkach, et al. Hexadecapolar colloids. Nat. Commun. \textbf{7}, 10659 (2016). \doi{10.1038/ncomms10659}
\bibitem{memory-araki} T. Araki,M. Buscaglia, T. Bellini, et al. Memory and topological frustration in nematic liquid crystals confined in porous materials. Nature Mater \textbf{10}, 303–309 (2011). \doi{10.1038/nmat2982}
\bibitem{memory-serra} F. Serra, M. Buscaglia, T. Bellini,
The emergence of memory in liquid crystals, Materials Today, \textbf{14} (10), 488-494, (2011). \doi{10.1016/S1369-7021(11)70213-9}
\bibitem{solitons3} G. Park, A. Suh, H. Zhao, C. Lee, Y.-S. Choi, I. I. Smalyukh, D. K. Yoon, Fabrication of Arrays of Topological Solitons in Patterned Chiral Liquid Crystals for Real-Time Observation of Morphogenesis. Adv. Mater. \textbf{34}, 2201749, (2022). \doi{10.1002/adma.202201749} 
\bibitem{solitons} G. Poy, A.J. Hess, A.J. Seracuse et al. Interaction and co-assembly of optical and topological solitons. Nat. Photon. \textbf{16}, 454–461 (2022). \doi{10.1038/s41566-022-01002-1}
\bibitem{solitons2} BX. Li, RL. Xiao, S. Paladugu et al. Three-dimensional solitary waves with electrically tunable direction of propagation in nematics. Nat .Commun. \textbf{10}, 3749 (2019). \doi{10.1038/s41467-019-11768-8}
\bibitem{tunabledefects} Sandford O’Neill, J.J., Salter, P.S., Booth, M.J. et al. Electrically-tunable positioning of topological defects in liquid crystals. Nat. Commun. \textbf{11}, 2203 (2020). \doi{10.1038/s41467-020-16059-1}
\bibitem{holestericstripes} L. Tran, M.O. Lavrentovich, D.A.
Beller, N. Li, K.J. Stebe, and R.D. Kamien, Lassoing Saddle-Splay and the Geometrical Control of
Topological Defects. Proc. Natl.
Acad. Sci. \textbf{113} 7106 (2016). \doi{10.1073/pnas.160270311}
\bibitem{confinement-organization} N. A. M. Araújo et al., Steering self-organisation through confinement, Soft Matter \textbf{19}, 1695–1704, 2023. \doi{10.1039/d2sm01562e}
\bibitem{confinement} D. Wang, M. Hermes, S. Najmr et al. Structural diversity in three-dimensional self-assembly of nanoplatelets by spherical confinement. Nat Commun \textbf{13}, 6001 (2022). \doi{10.1038/s41467-022-33616-y}
\bibitem{symmetry-breaking}  A. Piccardi, A. Alberucci, N. Kravets, O. Buchnev, G. Assanto, Nematicon-enhanced spontaneous symmetry breaking, Molecular Crystals and Liquid Crystals, 649:1, 59-65, (2017) \doi{10.1080/15421406.2017.1303916}
\bibitem{souslov2017} A. Souslov, B. van Zuiden, D. Bartolo, and V. Vitelli, Topological sound in active-liquid metamaterials, Nat. Phys. {\bf 13}, 1091 (2017). \doi{10.1038/nphys4193}
\bibitem{scheibner2020} C. Scheibner, A. Souslov, D. Banerjee, P. Sur\'owka, W.T.M. Irvine, and V. Vitelli, Odd elasticity, Nat. Phys. {\bf 16}, 475 (2020). \doi{10.1038/s41567-020-0795-y}
\end{thebibliography}
\end{document}